\documentclass[useAMS,usenatbib]{mn2e}
\usepackage{graphicx}
\usepackage{ifpdf}
\usepackage{epstopdf}
\usepackage{url}
\usepackage{multirow}
\usepackage{float}

\usepackage{mwe}

\usepackage{color}

\usepackage[dvipsnames]{xcolor}

\usepackage{amssymb}

\usepackage{pdflscape}

\usepackage{booktabs} 
          \usepackage{multirow}
          \usepackage{array}
          \usepackage{longtable}
           \usepackage{tabu}
            \usepackage{threeparttable} %table with notes
             \usepackage{threeparttablex} %longtable with notes
             \usepackage{rotating}
              \usepackage{xtab}
               \usepackage{multicol}

              \usepackage[title,titletoc,toc]{appendix} % \appendix

\usepackage{caption}

\usepackage{amsmath}
\let\oldAA\AA
\renewcommand{\AA}{\text{\normalfont\oldAA}}

\title[CCSNe ages and metallicities]{Core-collapse supernovae ages and metallicities from emission-line diagnostics of nearby stellar populations.}
\author[Lin Xiao, L. Galbany, J.J. Eldridge and Elizabeth R. Stanway]{Lin Xiao$^{1,2,3}$\thanks{E-mail: lxiao33@ustc.edu.cn}, L. Galbany$^{4}$\thanks{E-mail: llgalbany@pitt.edu}, J.J. Eldridge$^{3}$\thanks{E-mail: j.eldridge@auckland.ac.nz}, and Elizabeth R. Stanway$^{5}$\thanks{E-mail: e.r.stanway@warwick.ac.uk }\\
\\$^{1}$CAS Key Laboratory for Research in Galaxies and Cosmology, Department of Astronomy, University of Science and Technology of \\ China, Hefei, 230026, China,\\$^{2}$School of Astronomy and Space Sciences, University of Science and Technology of China, Hefei 230026, China, \\$^{3}$Department of Physics, University of Auckland, NZ,\\$^{4}$PITT PACC, Department of Physics and Astronomy, University of Pittsburgh, Pittsburgh, PA 15260, USA,\\$^{5}$Department of Physics, University of Warwick, Gibbet Hill Road, Coventry, CV4 7AL, UK}
\begin{document}

\pagerange{\pageref{firstpage}--\pageref{lastpage}} \pubyear{2017}
\maketitle

\label{firstpage}

\begin{abstract}
Massive stars are the main objects that illuminate H\,II regions and they evolve quickly to end their lives in core-collapse supernovae (CCSNe). Thus it is important to investigate the association between CCSNe and H\,II regions. In this paper, we present emission line diagnostics of the stellar populations around nearby CCSNe, that include their host H\,II regions, from the PMAS/PPAK Integral-field Supernova hosts COmpilation (PISCO). We then use BPASS stellar population models to determine the age, metallicity and gas parameters for H\,II regions associated with CCSNe, contrasting models that consider either single star evolution alone or incorporate interacting binaries. We find binary-star models, that allow for ionizing photon loss, provide a more realistic fit to the observed CCSN hosts with metallicities that are closer to those derived from the oxygen abundance in O3N2. We also find that type II and type Ibc SNe arise from progenitor stars of similar age, mostly from 7 to 45\,Myr, which corresponds to stars with masses ${\rm \leq 20M_{\odot} }$. However these two types SNe have little preference in their host environment metallicity measured by oxygen abundance or in progenitor initial mass. We note however that at lower metallicities supernovae are more likely to be of type II.\\
\end{abstract}


\section{Introduction}
Core-collapse supernovae (CCSNe) originate in massive stars ($ \geq 8 {\rm M_{\odot}} $), as a result of the gravitational collapse of the iron-group element cores that are the final result of their core nuclear burning. Given their short lifetimes, these massive CCSN progenitors tend to be associated with their birth place (star clusters or H\,II regions) with only a low possibility of being ejected from their birth place due to dynamical interactions or supernova kicks \citep{1997A&A...318..812D,2005A&A...437..247D,2011MNRAS.414.3501E,2018arXiv180409164R}. With direct detections of CCSN progenitor stars remaining rare \citep{2015PASA...32...16S}, the study of their environment represents a good alternative to put constraints on their characteristics \citep[e.g][]{2014ApJ...791..105W,2017A&A...601A..29Z,2018arXiv180308112W}. 

Here we focus on the emission line nebulae, or H\,II regions, associated with CCSNe. Previous studies of CCSN locations within their host galaxies have claimed various associations with H\,II regions. \cite{1992AJ....103.1788V} undertook the first attempt to assess the CCSNe association with H\,II regions based on a sample of 38 CCSNe of all subtypes, and concluded that approximately 50 per cent were associated with an H\,II region, with no statistically significant difference between type II (hydrogen rich) and Ibc (stripped-envelope) core collapse SNe. More recently, \cite{2008MNRAS.390.1527A} found a low fraction of type II SNe to be associated with H\,II regions while type Ibc SNe are spatially coincident with them, which led to an interpretation that the progenitors of type Ibc SNe are more massive than those of type II SNe. \cite{2013MNRAS.428.1927C} examined the immediate environments of 39 CCSNe in nearby galaxies and obtained similar results to \cite{2008MNRAS.390.1527A}, but argued that the observed association between certain CCSNe and H\,II regions provides only weak constraints upon their progenitor masses due to the fact that these CCSN hosts H\,II regions are long-lived giant H\,II regions ($ \sim $ 20\,Myr) rather than short-lived ($ \sim $ 4\,Myr) isolated, compact H\,II regions where most star-formation occurs. 

Other studies, for example \citet{2014ApJ...791..105W,2018arXiv180308112W}, have instead focused on the surrounding stellar populations of CCSNe. By studying the resolved stellar population around the site of a CCSN and assuming the age is similar to that of the progenitor, ages from a few Myrs up to a few $\times$10 Myrs have been found, in agreement with the range of ages possible for SNe from theoretical predictions \citep[e.g.][]{2017A&A...601A..29Z}. The ages and progenitor masses also therefore agree with those for the directly detected SN progenitors \citep{2015PASA...32...16S}. 

In addition to estimating the age of a SN progenitor, its metallicity is also important. The gas-phase metallicity can be estimated for the CCSN host environment from emission-line diagnostics. To accurately explain the differences in observed CCSN progenitors we must understand both their ages and metallicities. \cite{2003A&A...406..259P} use the metallicity-luminosity relation for late type galaxies and found the observed ratio of type Ibc to type II SNe depended strongly on the metallicity of the host galaxy. In those more luminous and metal-rich galaxies higher ratios of type Ibc to type II are expected. Similar results for the dependence of the relative ratio of SN subtypes on metallicity were obtained by many authors including \citet{2008ApJ...673..999P} and \citet{2017ApJ...837..120G} for example. We note that stellar evolution theory suggests that this relationship between ratio of SN types and metallicity can provide important constraints on how important stellar rotation and binary interactions are in the evolution of the progenitor stars \citep[e.g.][]{{1992ApJ...391..246P},1998A&A...333..557D,2003ApJ...591..288H,2004A&A...422..225M,2008MNRAS.384.1109E,2017A&A...601A..29Z}.

Lately, integral field spectroscopy (IFS) has begun to enable larger scale investigation over both spatial and spectral dimensions to investigate SN environments and active star-forming regions. For example, \cite{{2013AJ....146...30K},{2013AJ....146...31K}} used small-field (6\arcsec$\times$6\arcsec) IFS observations to identify single stellar clusters that had hosted CCSNe. They estimated their metallicity from the measured emission lines via strong-line methods, and their age by comparing H$\alpha$ emission equivalent width to simple stellar population models. They were able to estimate both the metallicity and initial mass of the CCSN progenitor stars and concluded that, on average, type Ibc SN explosion sites are more metal-rich than type II sites, and that some type II SNe progenitors may have been stars with masses comparable to those of type Ibc SN progenitors. 

Another advantage of IFS is that it enables simultaneous investigation of the overall properties of the host galaxy as well as its spatially resolved structure. This allows more constraints on the nature of the progenitors of different SN types, by looking for differences in local environmental parameters, as well as relating them to the overall distribution across the galaxy extent, as shown by \cite{{2014A&A...572A..38G},{2016A&A...591A..48G}}. These works constructed statistical samples to compare the star formation density and metallicity of stars at the locations of different SN types and confirmed that SN Ib/c show the closest relation to star-forming regions and are more associated with metal-rich environments than SN II. Similar results have been found in other recent studies based on different observed samples, such as those studying SN explosion sites with MUSE IFS \citep{2016MNRAS.455.4087G,2017arXiv171105765K} and PISCO, the largest updated sample of SN host galaxies observed with IFS which consists of 272 SNe including 152 CCSNe \citep{2018arXiv180201589G}.

However, all the above IFS studies assume one (mostly unstated) caveat. That is that above studies all assume stars evolve isolated as single stars.  In fact, over 70 per cent stars of massive stars are found in binary or multiple systems \citep[e.g][]{2012Sci...337..444S,2014ApJS..215...15S}. Binary interactions cause mass transfer between stars and therefore lead to new evolution pathways with respect to single star evolution and substantially change the appearance of stellar populations \citep[e.g][]{2017PASA...34...58E,2018MNRAS.477..904X}. In this work, we aim to investigate the effect of interacting binaries on SNe and the emission lines from the nearby host stellar populations. This is then used to constrain the ages and metallicities of the CCSN progenitor population taking full account of interacting binary stars.

To achieve our main aim to explore the effects of binary evolution on the nature of CCSN progenitor stars and their host environment we use the latest BPASS (Binary Population Spectral and Synthesis) models v2.1 \citep{2017PASA...34...58E} and our nebular emission line models that have previously been discussed in \cite{2018MNRAS.477..904X}. Using these we derive the properties of the underlying stellar population from the best-fitting models that match each individual H\,II regions. 

This paper is organized as follows. In Section 2, we describe the characteristics of the observed sample of CCSN host H\,II regions from PISCO, highlight some important observation quantities and discuss the photoionization map behaviour of these CCSN hosts. Then, in  Section 3, we briefly explain our numerical method for nebular emission models and the selection of best-fitting models which is discussed more in detail in \cite{2018MNRAS.477..904X}. In Section 4, we describe the best-fitting models in terms of their oxygen abundance comparing with observed value, as well as other physical parameters that determined the H\,II region models. Then in Section 5, we included the effect of ionizing photon leakage on best-fitting models and described the results in leakage case compared to those without leakage. Section 6 present further discussions on the uncertainties of our model and result. Finally, in Section 7, we summarize and give our conclusions.

\section{Sample of CCSN Host H\,II Regions and Dataset Analysis}
\begin{table}
\caption{The average value of oxygen abundance derived by strong-line method described in Section 2.2 and the average emission line ratios of the CCSN host H\,II regions.}

\begin{center}

                      \begin{tabular}{l@{\hskip 0.2in} @{\hskip 0.2in}c@{\hskip 0.2in}  @{\hskip 0.2in}c}
                        \hline
                        \hline
                        \rule{0pt}{3.4ex}
                        SN Type & SN II & SN Ibc \\ 
                        \hline 
                        12 + ${\rm \log(O/H) }$ & $ 8.45 \pm 0.10 $  & $ 8.49 \pm 0.10 $  \rule{0pt}{3.4ex}\\
                        EW(H$ {\rm \alpha} $)/$\AA$ & $ 50.9 \pm 49.1 $ & $ 49.9 \pm 42.0 $ \rule{0pt}{3.4ex}\\
                       $ {\rm \log([O\,{\sc III}]/H\beta) } $ & $ -0.19 \pm 0.32 $ & $ -0.29 \pm 0.33 $  \rule{0pt}{3.4ex}\\ 
                       $ {\rm \log([N\,{\sc II}]/H\alpha) }$  & $ -0.53 \pm 0.22 $ & $ -0.50 \pm 0.20 $  \rule{0pt}{3.4ex}\\
                       $ {\rm \log(S\,{\sc II}/H\alpha) }$  & $ -0.45 \pm 0.13 $ & $ -0.51 \pm 0.11 $  \rule{0pt}{3.4ex}\\
                       $ {\rm \log(O\,{\sc I}/H\alpha) }$   & $ -1.28 \pm 0.25 $ & $ -1.45 \pm 0.19 $  \rule{0pt}{3.4ex}\\
                        \hline
                        \hline
                        
\end{tabular}

\end{center}
\label{tab:BPT_ave}
\end{table}

\subsection{Overview of the Sample}

The sample of CCSN host H\,II regions used in this work comes from the PMAS/PPAK Integral-field Supernova hosts COmpilation \citep[PISCO,][]{2018arXiv180201589G}. The PISCO compilation started as an extension of the CALIFA survey targeting low mass SN host galaxies that were missing in the CALIFA mother sample. So the instrumental configuration, observations and reduction is performed following CALIFA procedures and reduction pipeline. All these information is well established and available following \cite{2018arXiv180201589G} and references there in, as well as the third CALIFA data release \citep[][and reference there in]{2016A&A...594A..36S}, including sky subtraction and flux calibration. Overall, it is composed of observations of CCSN host galaxies selected from the CALIFA survey and other dedicated programs using the PMAS/PPAK integral field unit with large field-of-view ($\sim$ 1 $\times$ 1 sq\,arcmin). The SN position was determined from the astrometry of the PISCO/CALIFA datacubes, and the aperture extraction was centred at that location. We required emission lines to have S/N $>$ 3 to be considered reliable, although in practice all have S/N $>$ 5. The spectral information covers most of the optical domain from 3750 to 7300\AA, with a spectral resolution ranging from $ \sim 2.7\AA $ in the blue to $ \sim 6\AA $ in the red. 

The observational sample we use consists of 152 CCSNe with 107 type II SNe and 45 type Ibc SNe, and their observed flux is derived within 1${\rm kpc^{2}}$ centred at the SN location. This is to provide us with adequate signal-to-noise in our spectra, while limiting the observation to the likely host stellar population around the SN site and preventing the varying distance to the host galaxy from causing artefacts in our analysis. Extragalactic HII regions as well as SN host H II clumps have typical diameters $\sim10-100$\,pc but range up to several hundred parsecs with the expansion of HII regions driven by nuclear activities and stellar feedback \citep{1997RMxAC...6..264G,2011ApJ...731...91L}. This implies that between 1 and 6 HII regions are included in a 1\,kpc$^2$ box \citep{2014A&A...561A.129M}, with relatively little dilution of the host region emission by unassociated nebular gas. Any contamination that is present will tend to be dominated by the youngest stellar population in the region and so bias CCSN host estimates to younger ages.The physical spatial resolution in the PISCO dataset is presented in the Figure 2 of \cite{2018arXiv180201589G}. The average resolution of CCSN host observations in this work is 300\,pc/arcsec, so on average we integrated the spectra within the area covered by 3x3 spaxels. Only complete spaxels are included (i.e. no correction is made for partial spaxels). Furthermore, we have investigated using different definitions for the ``nearby host stellar populations", including using the emission-line strengths from the nearest H\,II region and from a square arcsecond around the SN location. We find that this choice does not affect our results to a large degree as shown in Appendix~\ref{sec:CCSNeAges}. Again, using the observed flux within 1 ${\rm kpc^{2}}$ allows fair comparisons among SN local environments because the region compared does not vary in each galaxy. If an angular size was used the region integrated would vary as the area would vary with the redshift of each galaxy. 

We also notice that all these CCSN host regions have H${\rm \alpha}$ and ${\rm [N\,{\sc II}]\lambda 6583 }$ emission while some may lack other emission lines, for example 1 of 152 CCSN hosts without H${\rm \beta}$ lines (SN 1989R), 9 of 152 hosts without ${\rm [O\,{\sc III}]\lambda 5007 }$, 8 of 152 hosts without ${\rm [S\,{\sc II}]\lambda 6713 }$ and 24 of 152 hosts without ${\rm [O\,{\sc I}]\lambda 6300 }$. But none of the hosts lose all these 4 lines, and most of them have only one emission line lost. Failure to detect individual emission lines usually indicates either that the line was coincidence with sky emission or absorption features, or that nebular emission from the host region is weak, and may be dominated by the underlying stellar population rather than the host environment. While, in principle, this means our survey may be biased against older HII regions, the effects on the distributions discussed in this paper are likely to be small. Approximately 20 per cent of type II CCSN locations identified in PISCO, and 15 per cent of type Ibc regions have at least one line undetected, but for the bulk of the population only one line is missing. As a result, while older populations may be missed, again potentially biasing the sample to younger ages, the nebular regions should be indicative of the range of properties in the CCSNe progenitor stellar population.

\subsection{Oxygen abundance}

\begin{figure}
\centering
\hspace*{-0cm}\includegraphics[width=9cm]{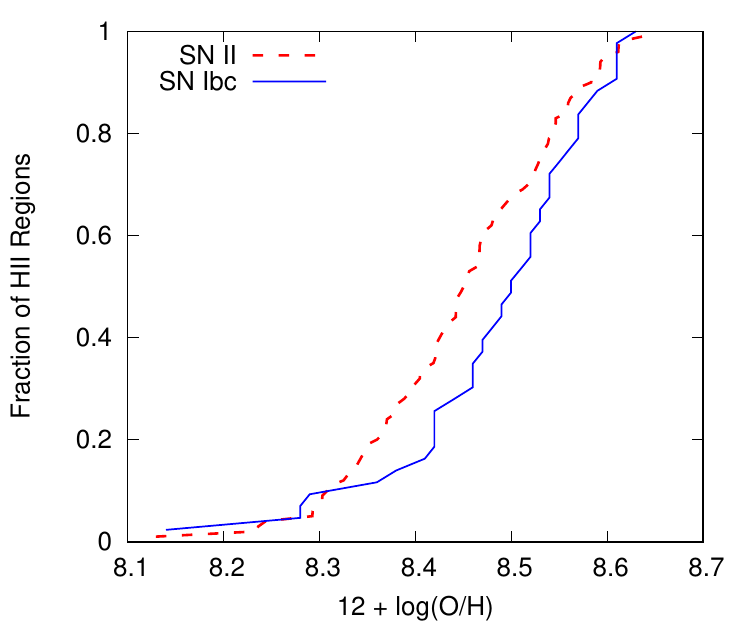}  % Frac_OH_1kpc.pdf
\centering
\caption[Cumulative distributions of the oxygen abundance of the CCSN host H\,II regions]{Cumulative distributions of the oxygen abundance of the CCSN host H\,II regions. The red dashed line is for type II SN hosts and blue solid line for type Ibc SN hosts.} \label{fig:CD_OH}
\end{figure}

As in many previous studies, we derive the gas-phase metallicity of the CCSN host H\,II regions from their measured fluxes via the strong-line method. This method is strongly affected by the choice of which strong-line abundance calibrations are used as discussed in \citet{2008ApJ...681.1183K} and \citet{2018MNRAS.477..904X}.  In this work, we use the most widely used empirical O3N2 calibrator as our standard metallicity calibration which is updated by \cite{2013A&A...559A.114M} using new direct abundance measurement (${\rm T_e}$-based) provided by \cite{2010ApJ...720.1738P}: 
\begin{equation}
{\rm O3N2} = \log {\rm([O\,{\sc III}]\lambda 5007/ H\beta)/([N\,{\sc II}]\lambda 6583/ H\alpha)},
\end{equation}
\begin{equation}
12 + \log({\rm O/H}) = 8.533 - 0.214 \times {\rm O3N2}.
\end{equation}

The new calibration was tested by comparison to the measured oxygen abundance of 3423 observed CALIFA H\,II regions using the multiple line-ratio calibrations \citep[][]{2010ApJ...720.1738P}. This improved O3N2 calibration shows weaker metallicity dependence and a relatively high precision with respect to other abundance determinations. This method was used in \cite{2016MNRAS.455.4087G} and \cite{2017arXiv171105765K} with MUSE data for the emission-line diagnostics of 11 and 83 SN host H\,II regions, respectively. %However, there are still some uncertainties between these indirect metallicity measurements with respect to theoretical methods as discussed by \cite{2018MNRAS.477..904X}. 

%The average values for emission line ratios, oxygen abundance and EW(H$ {\rm \alpha} $) of the sample are listed in Table. \ref{tab:BPT_ave}, which shows that type II and type Ibc SN hosts have very similar average oxygen abundance. 
Figure~\ref{fig:CD_OH} demonstrates the cumulative distribution of the oxygen abundances of type II and type Ibc SNe separately. The results indicate that these two types of CCSNe span in a similar oxygen abundance range from 8.1 to 8.7 and the main fraction (about 80 per cent) is located in oxygen abundance from 8.3 to 8.6. Due to the steeper increasing trend of type Ibc SNe, their overall metallicity is higher than that of type II SNe. This result is reflected in their average oxygen abundance given in Table \ref{tab:BPT_ave}. Type Ibc have average of 8.49, slightly higher than type II of average of 8.45.

\subsection{Equivalent Width of H$ \alpha $}
\begin{figure}
\centering
\hspace*{-0cm}\includegraphics[width=9cm]{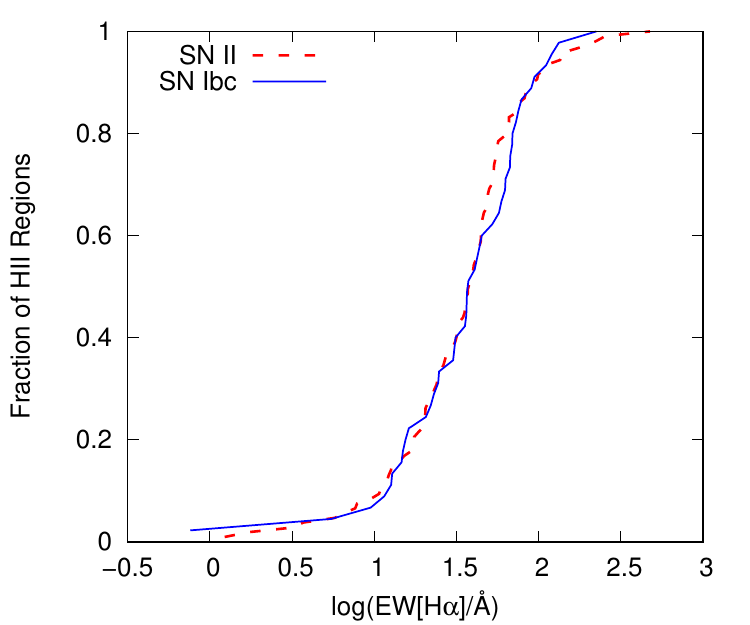}  % Frac_EW_1kpc.pdf 
\centering
\caption{The cumulative distribution of CCSN hosts H\,II regions as a function of H$ \alpha $ EW with SNe II in red dashed line and SNe Ibc in blue line. } \label{fig:CD_EW}
\end{figure}

The H$ \alpha $ line luminosity is widely used to measure the amount of ongoing star formation. The H$ \alpha $ equivalent width (EW), the measure of the relative strength of the line with respect to the continuum, is commonly used to infer age of a starburst. This is reasonable as the strength of this particular line is largely dependent on the ionizing condition of emission nebulae which is dominated by the short-lived massive stars and the continuum light is dominated by the long-lived lower-mass stars. Thus, the H$ \alpha $ EW varies with age and star-formation history of the galaxy. 

Figure~\ref{fig:CD_EW} shows the cumulative distribution of H$ \alpha $ EW for our SN hosts sample, where the two SN type hosts have a similar distribution with nearly 90 per cent of SNe having H$ \alpha $ EW below 100\AA~and above 8\AA~which indicates most SN host regions have significant hydrogen recombination lines from the nearby stellar populations. These similarity of the distributions is also shown by their almost identical mean H$ \alpha $ EW as seen in Table \ref{tab:BPT_ave}. Surprisingly a type II SN host has the highest H$ \alpha $ EW of up to 400\AA, compared to that of the type Ibc SNe of 250\AA. There are many reasons that this might be, but it does indicate that while EW is a useful estimator there are many factors that limit its accuracy in understanding the age of SN progenitors.

An important question in this study is whether the SNe are associated with a HII region or not. The H$ \alpha $ EW distribution also allows us consider this question. The typical H$\alpha$ EW of a HII region varies but we see in our sample that there is a smooth distribution down to a EW of about 10\AA \,below which the distribution flattens out. This is perhaps indicative that all stellar populations have some associated nebular emission but only the brightest and most luminous are identified as HII regions. We do not attempt to determine the dividing line. If for example we were to say a EW of 100\AA \,and above indicated a HII region then very few of our SNe would be associated with such a region. Most of our SNe are associated with nebular emission that is typically a few times 10\AA \,in strength.

\begin{figure}
\centering
\vspace*{-0.0cm}
\hspace*{-0cm}\includegraphics[width=9cm]{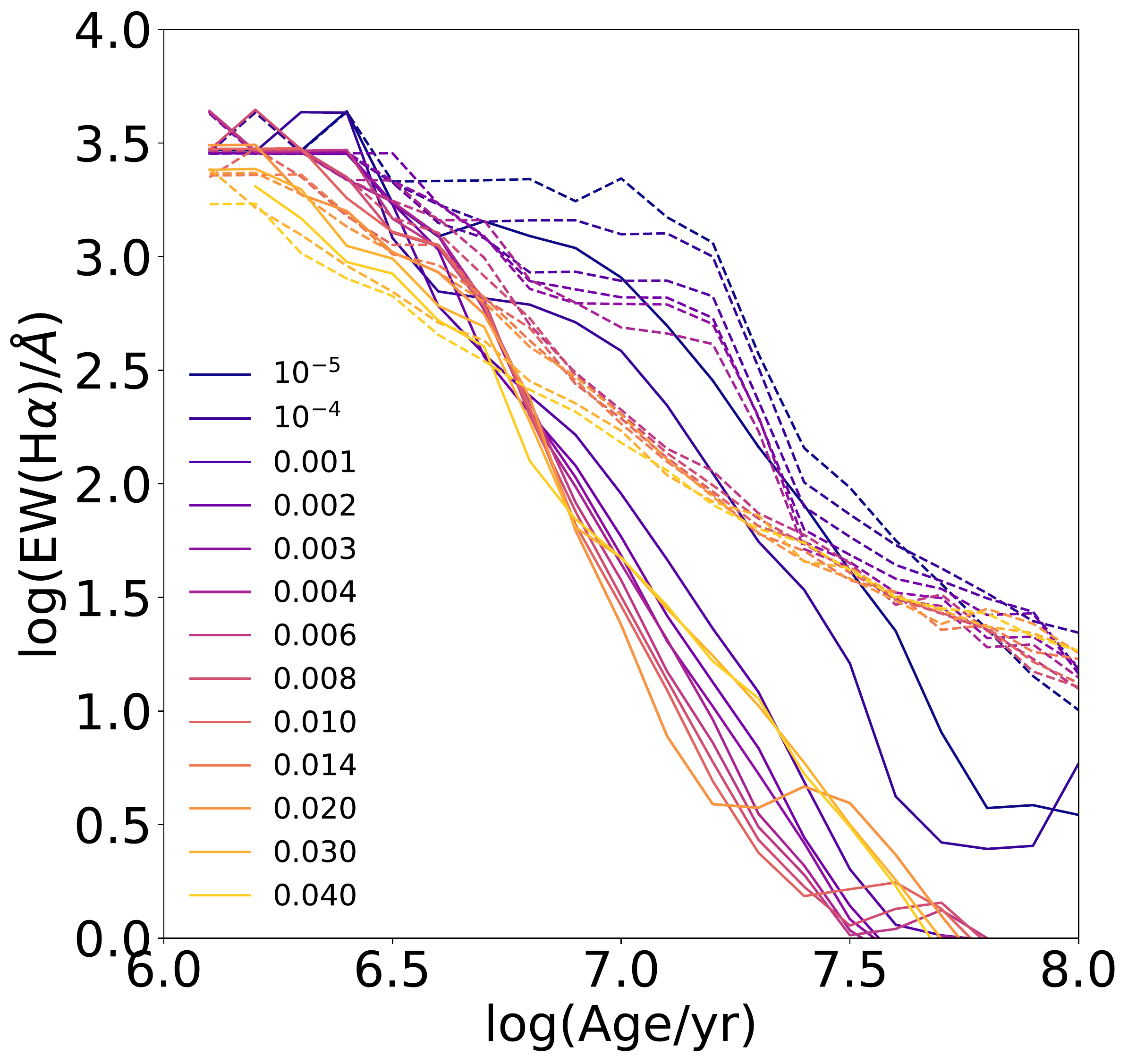}  % EW_Ha-BW.pdf / EW_Ha.pdf
\centering
\caption[H$ \alpha $ EW as a function of age at ${\rm \log(n_H/cm^3)=2.0 }$ and ${\rm \log(U)=-2.5 }$ for both single-star and binary-star models]{H${\rm \alpha }$ EW as a function of age at ${\rm \log(n_H/cm^3)=2.0 }$ and ${\rm \log(U)=-2.5 }$ for both single-star (solid lines) and binary-star models (dashed lines) at increasing metallicities represented from blue to yellow. } \label{fig:EW_Ha}
\end{figure}

In addition, we have tested how interacting binaries might affect the accuracy of age estimates from H$\alpha$ EW. To do this, we use BPASS models v2.1 emission nebulae models \citep{2017PASA...34...58E,2018MNRAS.477..904X} to calculate the H$ \alpha $ EW as a function of age and metallicity using only single stars or binary stars in the stellar population forming the ionizing source, as shown in Figure~ \ref{fig:EW_Ha}. We here assume a constant hydrogen density of ${\rm \log(n_H/cm^3)=2.0 }$ and ionization parameter ${\rm \log(U)=-2.5 }$. 

For the first 3\,Myr, both single-star and binary-star models have a high value of H$ {\rm \alpha} $ EW around 3000\AA. Afterwards the H$ {\rm \alpha} $ EW of single-star models experience an abrupt decrease to less than 1\AA\ in 30\,Myr. In contrast, binary-star models have a slower H$ {\rm \alpha} $ EW decline and have an EW of around 30\AA\, at the same age. This is because binary interactions strongly enhance the emission lines' strength at  ages beyond 10\,Myr producing a clear difference to single stars, due to the hot WR or helium stars produced at later times via binary interactions \citep[e.g.][]{1999NewA....4..173V,2017A&A...608A..11G,2018MNRAS.477..904X}

The H$ {\rm \alpha} $ EW is also strongly affected by metallicity as the main-sequence lifetime of massive stars is slightly increased at lower metallicity and stars are more compact and thus hotter. This leads to the low-metallicity model EW decreasing slower than those at higher metallicity. In addition, binary-star models naturally separate into two groups: a low-metallicity group of Z $ \leq $ 0.004 that have a long plateau phase up to 20\,Myr, and high-metallicity group of Z $ \geq $ 0.006 which gradually decreases without the plateau. The plateau is due to inclusion of quasi-homogeneous evolution in low metallicity binary-star systems where mass-transfer has occurred. For stars with metallicities Z $ \leq $ 0.004 and initial masses above 20M$_{\odot}$ that accrete more than 5 per cent of their initial mass we assumed to be spun up so rapidly they evolve fully mixed over their main sequence lifetimes. This is discussed in detail in \citet{2011MNRAS.414.3501E} and \citet{2017PASA...34...58E}. The mixing significantly extends the lifetime of massive stars in lower metallicity models and they avoid a cool red supergiant phase and thus emit more ionizing photons than they would otherwise. In contrast, at metallicities above Z$>$0.004 H$ {\rm \alpha} $ EW is only weakly dependant on metallicity. Thus given the moderate metallicities of our sample as shown in Figure \ref{fig:CD_OH} the shape of our observed sample distribution in Figure \ref{fig:CD_EW} is not strongly affected by metallicity.

\begin{figure}
\centering
%\vspace*{-0.5cm}
\hspace*{-0cm}\includegraphics[width=8.5cm]{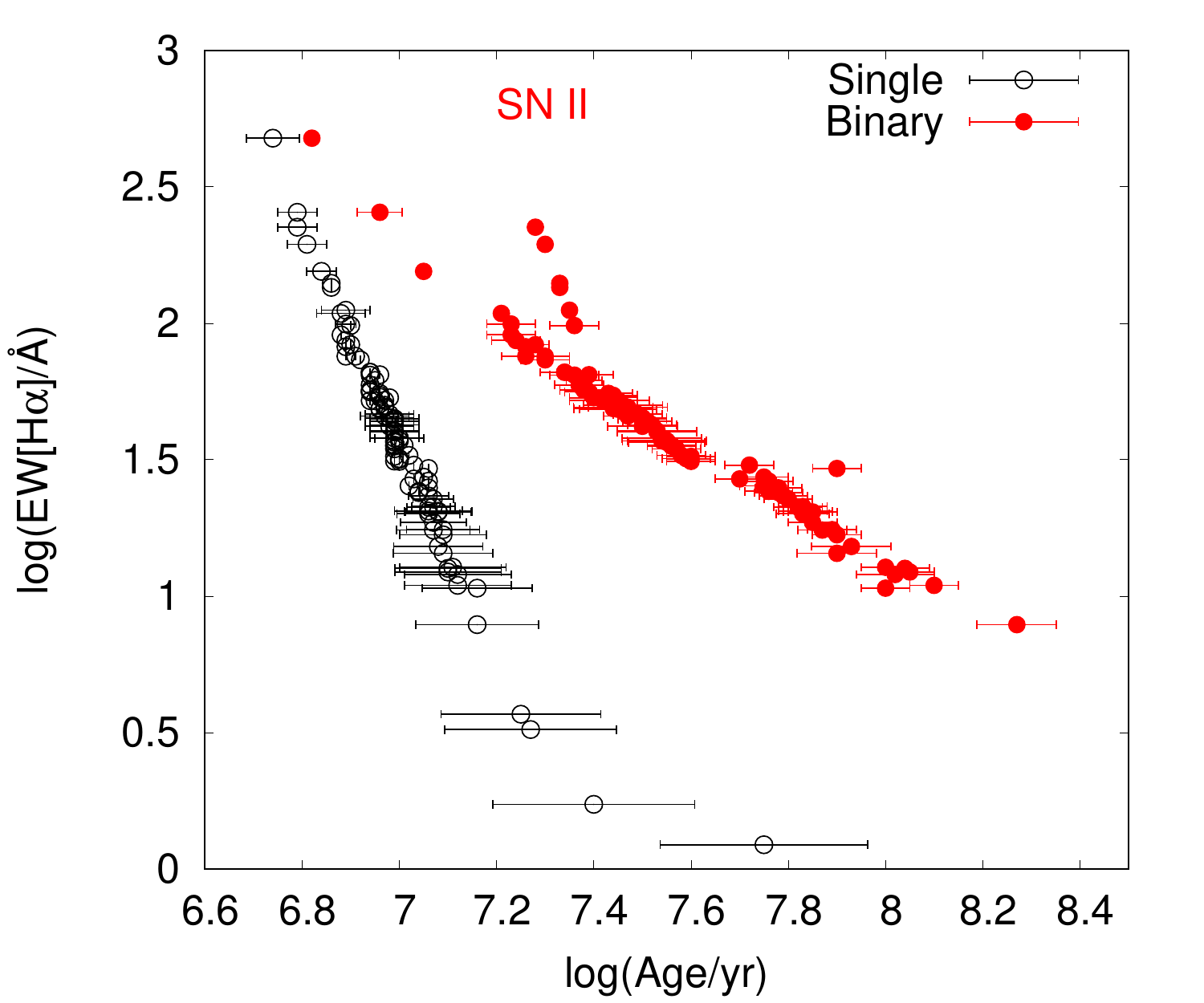}\\ % SNII_age_1kpc.pdf
\hspace*{-0cm}\includegraphics[width=8.5cm]{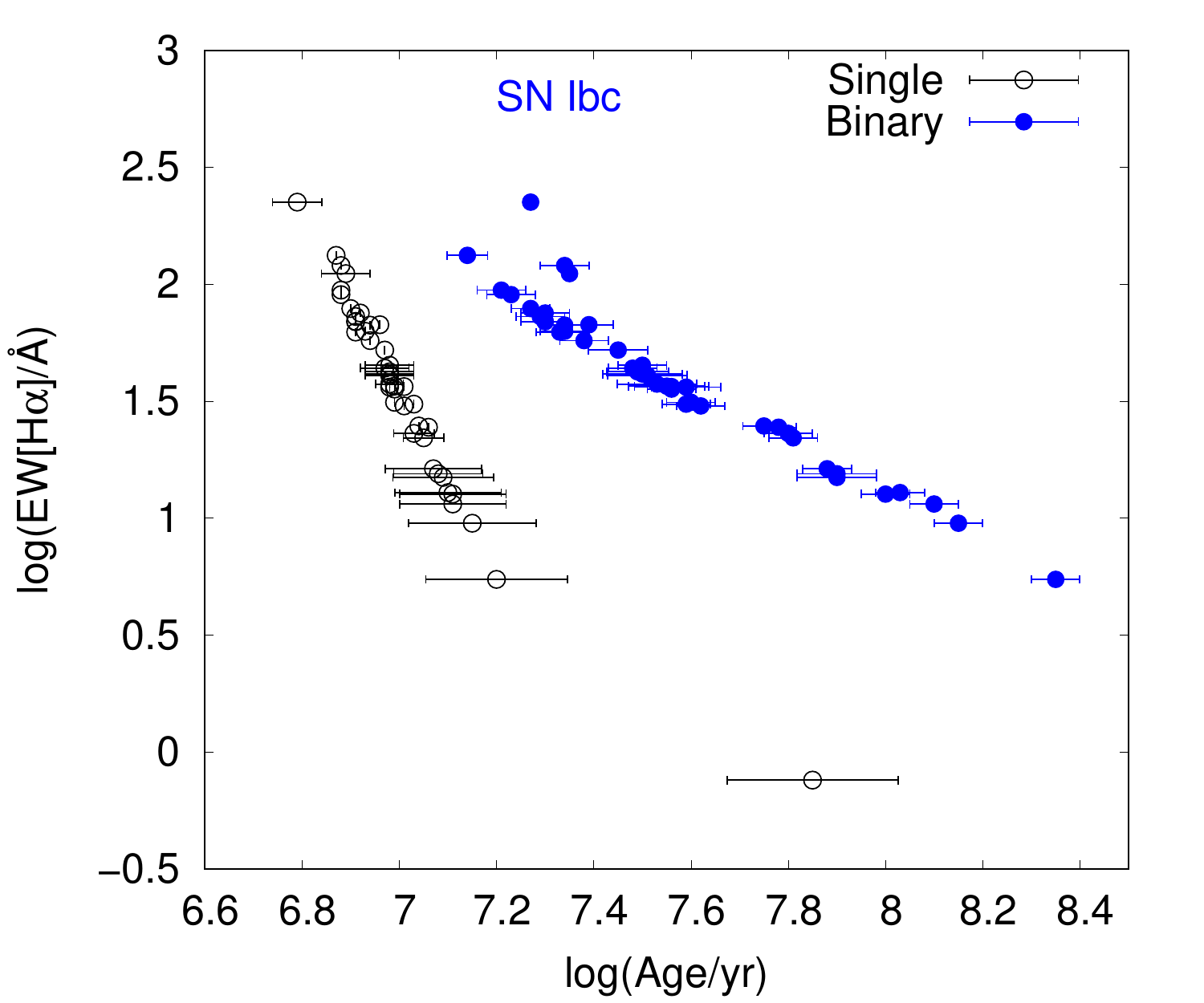}  % SNIbc_age_1kpc.pdf
\caption[The distribution of H$ {\rm \alpha} $ EW with respect to age for type II and type Ibc SNe.]{The distribution of H$ {\rm \alpha} $ EW with respect to age for type II (the top panel) and type Ibc (the bottom panel) SNe, assuming the simple relations described in section 2.3. The black circles with error bars are for single-star models and red (SN II) and blue (SN Ibc) dots with error bars for binary-star models. The errors are due to the uncertainty of observed H$ {\rm \alpha} $ EW values and the variance of ionization parameter from ${\rm \log(U)} = $ -3.5 to -1.5, at ${\rm \log(n_H/cm^3)=2.0 }$. } \label{fig:SN_age}
\end{figure}

Using our BPASS EW-age relationship and the metallicity derived using O3N2 calibration, we estimate the age separately for SN II and SN Ibc from the H$ {\rm \alpha} $ as shown in Figure~\ref{fig:SN_age}. As the H$ {\rm \alpha} $ EW decreases with increasing age we are able to simply estimate the age for all SN hosts. Using only our single-star models most SNe have a progenitor age $ \leq $ 20\,Myr, while using the binary models leads to most ages being above this maximum with ages up to 200\,Myrs being possible. We can conclude that binary interactions can significantly extend CCSN progenitor ages estimated from H$\alpha$ EW. Although even when using the interacting binary models we note this method of estimating the age is highly uncertain due to the large scatter in the metallicity calibration, the sensitivity of the line strength to assumed metallicity, and also the uncertainties associated with the assumed binary fraction. Therefore, this age estimation should only be considered as an upper limit ($ \gtrsim $50 Myr) becuase there are no Galactic nebular regions that we have calibrated and verified models of such extreme ages against. Although we have performed some validation of such old models in \cite{2018MNRAS.477..904X}. In addition, considering that the CCSNe host in old and diffused environment, the measured H$ {\rm \alpha} $ EW can be more contributed by the underneath stellar populations rather than the emission nebulae. Therefore, these estimated ages can be longer than the typical lifetime of HII regions (up to a few 10 Myrs), and this is consistent with the CCSN ages of \cite{2017MNRAS.471.1390H} using completely different approach based on the CCSN spatial distribution in their host galaxies. While in the real galaxies, planetary nebulae and other sources from x-ray binaries \citep{2016MNRAS.455.1770W} can get into these late ages that are not yet in BPASS models.

\subsection{Distribution of CCSN Hosts in BPT Diagrams}

\begin{figure*}
\centering
\hspace*{-1.0cm}\includegraphics[width=18cm]{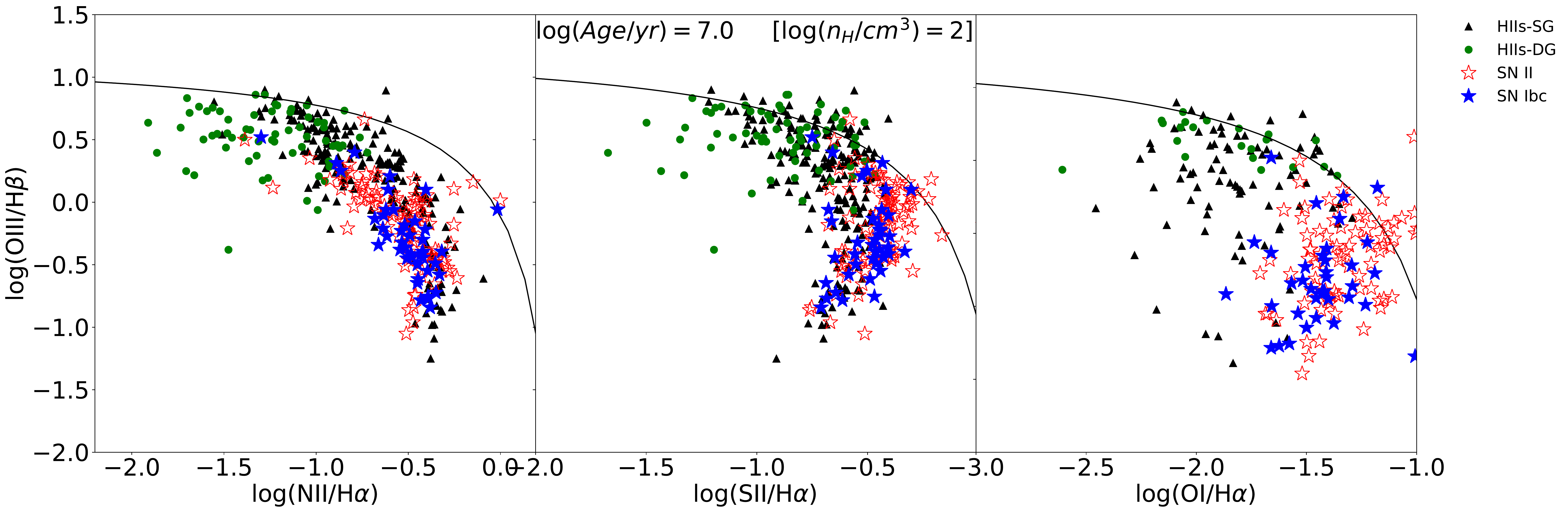}  % BPT_SNe_1kpc.png
\centering
\caption[Distributions of CCSN Host H\,II Regions in BPT diagrams]{Distributions of CCSN host H\,II regions in BPT diagrams compared to van Zee H\,II region sample (the black triangles from spiral galaxies and green circles from dwarf galaxies. The type II SN hosts are represented by the red stars and type Ibc SN hosts are represented by the blue stars. } \label{fig:BPT_SNe}
\end{figure*}

\begin{figure*}
\centering
%\vspace*{-0.5cm}
\includegraphics[width=18.cm]{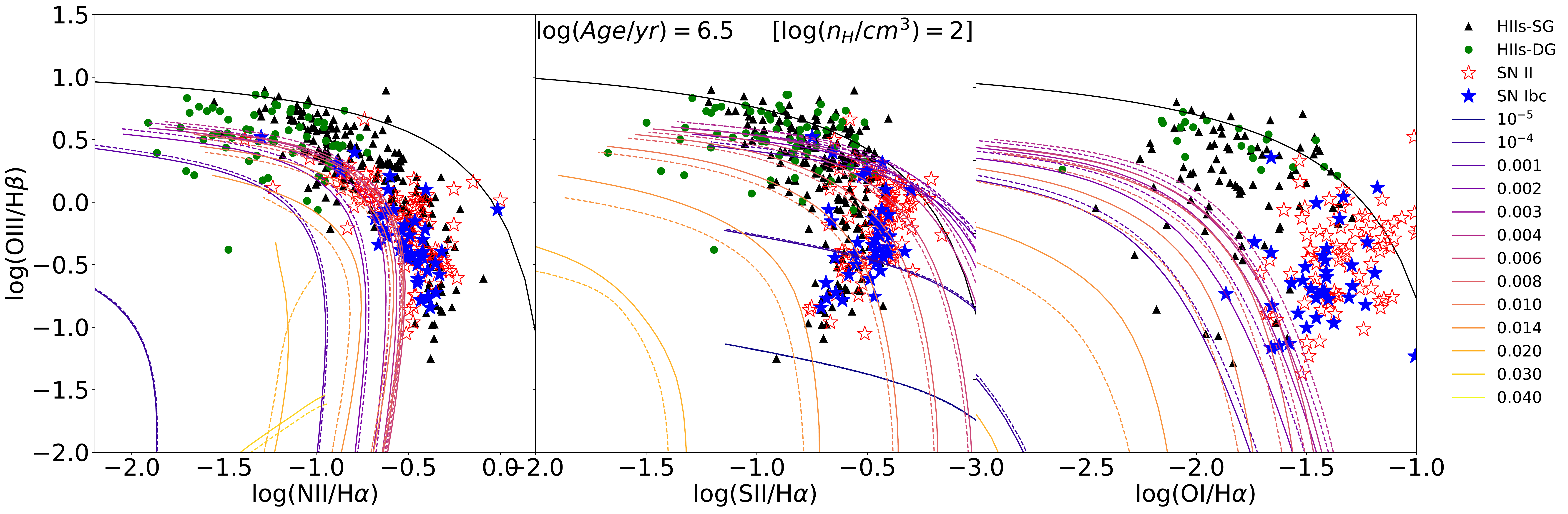}\\ % BPT6p5_Hd2_sb_1kpc.pdf
\includegraphics[width=18.cm]{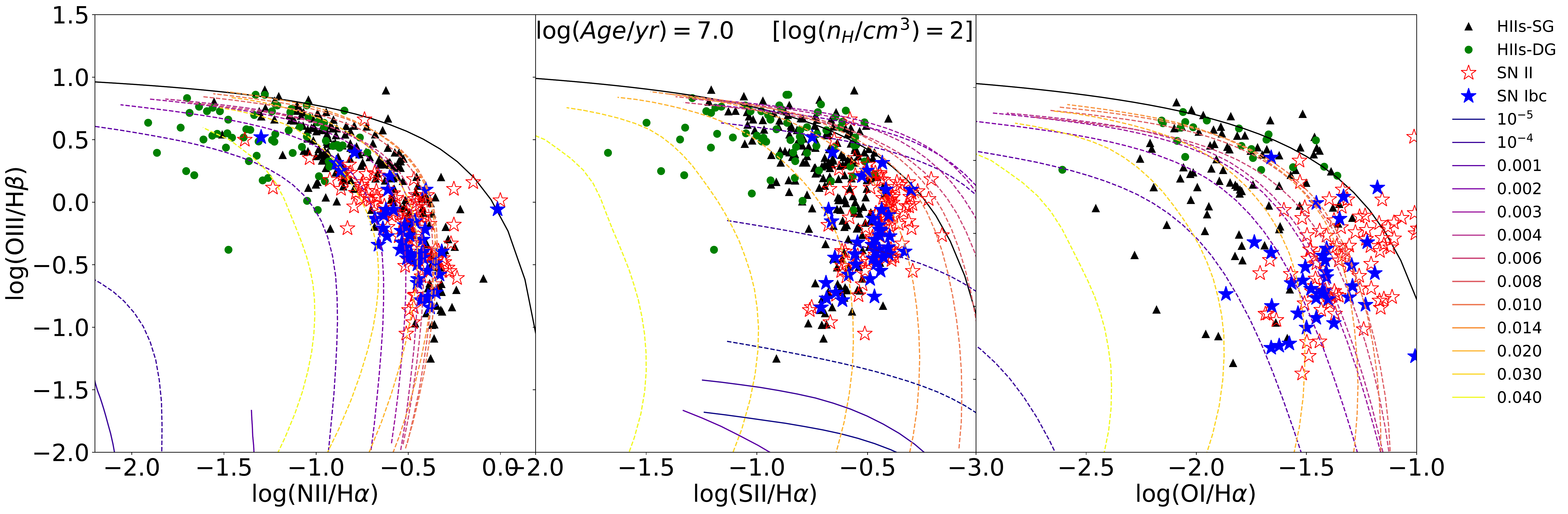}   % BPT7p0_Hd2_sb_1kpc.pdf
\centering
\caption[The BPT diagrams of \textsc{bpass} models at 3 and10 Myr]{The BPT diagrams of \textsc{bpass} models at 3\,Myr in the top panel and 10\,Myr in the bottom panel. The tracks in solid lines are from single-star models and those in dashed lines are binary-star models. The tracks with different colours stands for different metallicities from blue to red metallicity growing from ${\rm Z = 0.00001 }$ to 0.040. The values of ionization parameter of the models is reduced following the track from upper left (${\rm \log(U) = -1.5 }$) to lower right (${\rm \log(U) = -4.5 }$).} \label{fig:BPT_SNe_BPASS}
\end{figure*}

As in \cite{2018MNRAS.477..904X}, we use the BPT diagrams first proposed by \cite{1981PASP...93..817B} as a basis to study the ionizing conditions of these CCSN host stellar populations and thus refine our estimates of age and metallicity. The distributions of the emission lines from the CCSN host stellar population are depicted in Figure~\ref{fig:BPT_SNe}. We compare them with the H\,II regions taken from \citet{1998AJ....116.2805V} and \citet{2006ApJ...636..214V}. These were H\,II regions identified in nearby galaxies. 

The majority of sources in both the van Zee sample of galaxy H\,II regions and our CCSN-host region sample lie beneath the maximum theoretical starburst lines defined by \cite{2001ApJ...556..121K}, although some scatter above these lines, particularly in the [S\,II] and [O\,I]-based diagnostics. The two samples do differ significantly with the H\,II regions of dwarf galaxies occupying the upper left region in the BPT diagram. The CCSN hosts are mostly in the lower right of the diagram indicating they are diffuse older nebula emission regions, without many stars hot enough to provide higher ${\rm [O\,{\sc III}]}$ to H$ \beta $ ratios. In addition, the CCSN hosts in the ${\rm [O\,{\sc I}]}$/H$ \alpha $ panel are more scattered left over the Kewley curve. Type II SN regions are more likely to exceed the limiting Kewley relation in ${\rm [O\,{\sc I}]}$/H$ \alpha $ than type Ibc hosts. Aside from this, the range of line ratios spanned by the two samples is very similar. They form a narrow sequence in the map of [OIII]/H$ \beta $ with [NII]/H$ \alpha $ as well as the [SII]/H$ \alpha $ diagram. Even though these hosts are dispersed across the map of [OIII]/H$ \beta $ with [OI]/H$ \alpha $, SNe II hosts and SNe Ibc hosts still occupy nearly the same regions. These similar behaviours of SN hosts in BPT diagrams can also be reflected from their similar mean emission line ratios as listed in Table~\ref{tab:BPT_ave}, where type Ibc SNe have slightly lower ratios in $ {\rm \log([O\,{\sc III}]/H\beta) } $, $ {\rm \log([S\,{\sc II}]/H\alpha) } $ and $ {\rm \log([O\,{\sc I}]/H\alpha) } $, but higher ratio of $ {\rm \log([N\,{\sc II}]/H\alpha) } $. %This difference could be due to the incompleteness of the observation dataset with fewer type Ibc than SN II SNe. 

\section{Method}

To model the nebular emission from stellar populations we follow the methods described in \cite{2018MNRAS.477..904X}. A short summary of our method is that we use the latest BPASS (Binary Population Spectral and Synthesis) models\footnote{The models are available at: \texttt{http://bpass.auckland.ac.nz}.} v2.1 \citep{2017PASA...34...58E} to obtain the input ionizing spectra for \textsc{CLOUDY} 13.03 \citep{1998PASP..110..761F,2013RMxAA..49..137F} nebular models.  We then use the \textsc{CLOUDY} output models to work out our emission line fluxes and equivalent widths. In addition to varying the input stellar age, initial metallicity and inclusion of interacting binaries we also vary the properties of the surrounding gas by varying the gas density and the ionization parameter. 

To aid our interpretation of these CCSN host nebula regions, we use our nebular emission models from \cite{2018MNRAS.477..904X} to predict where stellar populations should lie in these BPT diagrams. Figure~\ref{fig:BPT_SNe_BPASS} shows a selection of our models at ages of $ {\rm \log(Age/yr) = 6.5 } $ \& 7.0, with a hydrogen density of ${\rm \log(n_{H}/cm^{3}) = 2 }$ over a reasonable range of ionization parameters.

As in \cite{2018MNRAS.477..904X}, the nebular emission models of different metallicities produced 13 separated tracks. The models with lower metallicities (${\rm Z < 0.020 }$), extend from upper left to lower right as the ionization parameter decreases from ${\rm \log(U) = -1.5} $ to -4.5. The three models with the highest metallicities (Z = 0.020, 0.030, 0.040) have a more vertical pathways that fall off with  decreasing ionization parameter. At 3\,Myr, neither models with too low nor too high metallicity can match the observed CCSN hosts, and models of ${\rm Z=0.001 }$ to 0.020 can only go through part of the CCSN host area. But as the population ages to 10\,Myr, tracks from single-star models die away quickly. In contrast, those from binary-star models move up and match all the nebula regions. This result again highlights the importance of binary interactions in interpreting the emission lines from these CCSN hosts at later times.

With our full suite of models we are able to perform a maximum likelihood fitting method to derive the preferred model parameters for each observed fitting region, rather than just comparing our models to the observed nebula regions by eye. For full details of the BPASS models, the nebular emission models and the fitting method we refer the reader to \cite{2017PASA...34...58E} and \cite{2018MNRAS.477..904X}. Using these nebular emission line models we derived the properties of the underlying stellar population from the best-fitting models that match each individual H\,II region, and the best-fitting parameters are listed in Tables~\ref{tab:bestfits} and \ref{tab:bestfits-leak}. We first use models that assume that none of the ionizing photon from the host stellar population are lost. Then second, we will allow for some ionizing photons leakage or loss. 

\begin{figure*}
\centering
\includegraphics[width=8.8cm]{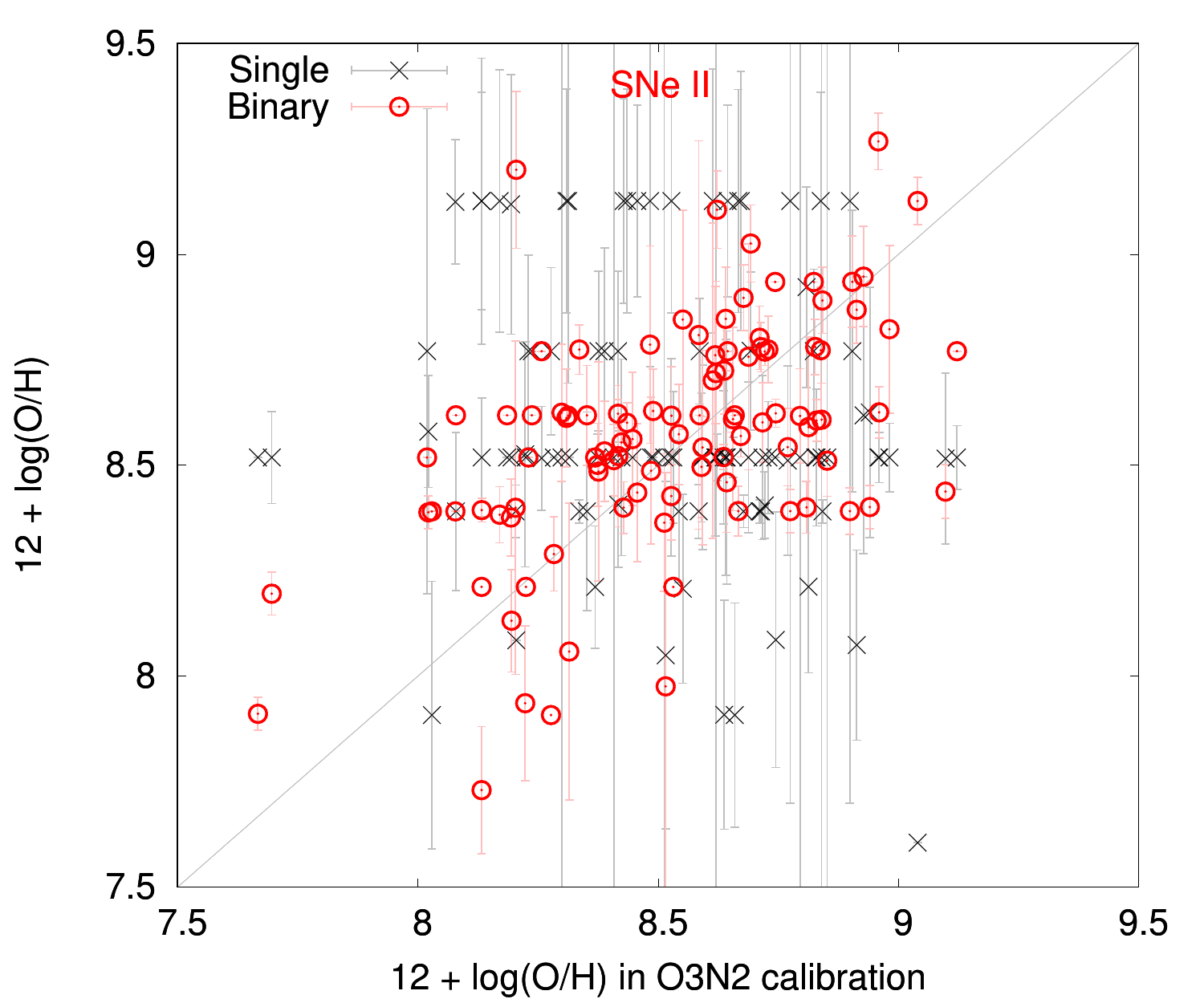}
\includegraphics[width=8.8cm]{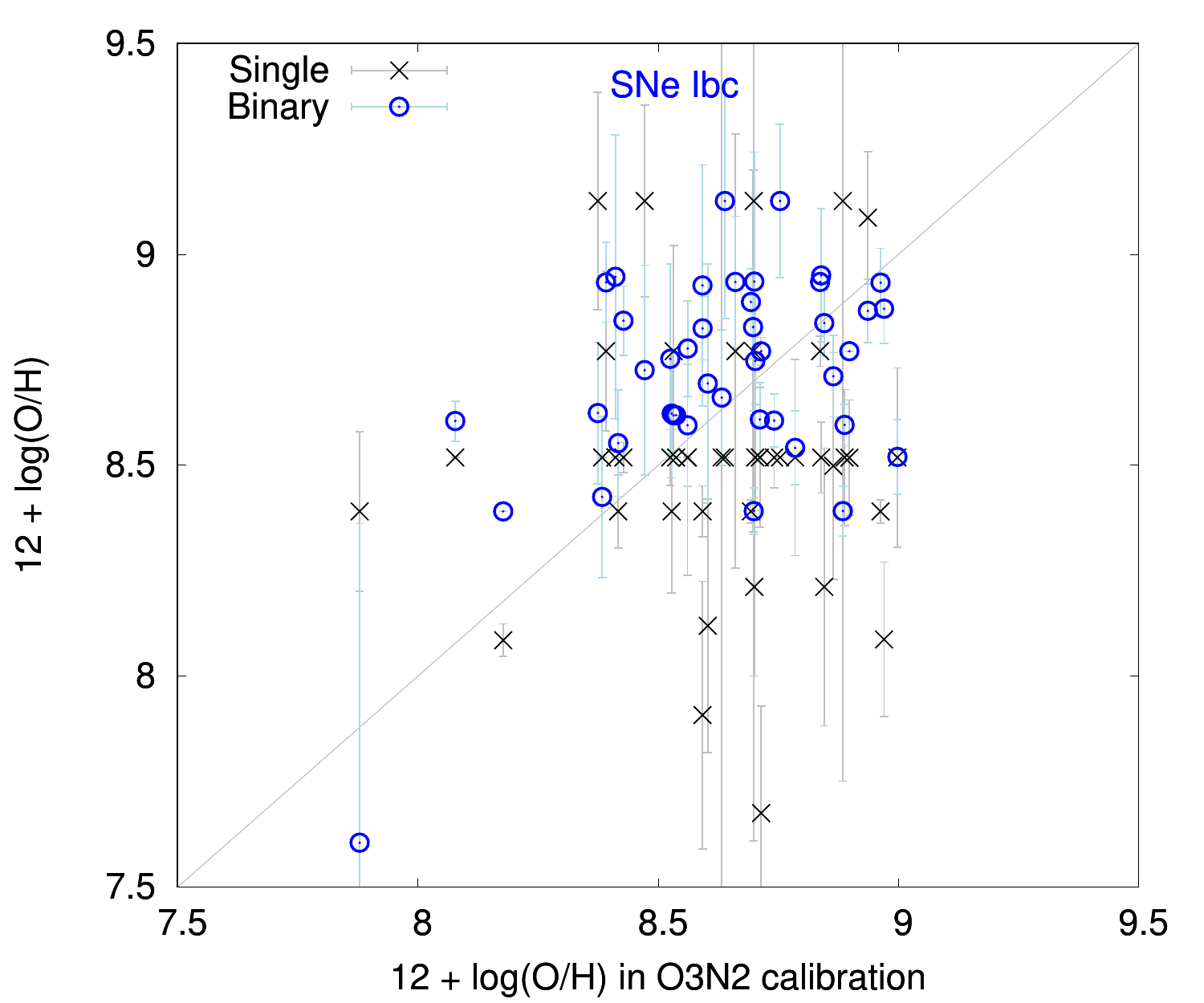}
\centering
\caption[The predicted oxygen abundance from best-fitting models comparing with those using O3N2 calibration]{Predicted oxygen abundance from best-fitting models comparing with those using O3N2 calibration. It shows the comparison between single- and binary-star models separately for type II SN hosts in the left and type Ibc SN hosts in the right. Single-star models are presented by the black crosses with error bars and binary models are the red (SNe II) and blue (SNe Ibc) circles with error bars.} \label{fig:OH_O3N2}
\end{figure*}

\begin{table*}
\caption{The average value and standard deviation of oxygen abundance 12 + ${\rm \log(O/H)}$, ionization parameter ${\rm \log(U)}$, hydrogen density ${\rm \log(n_H/cm^3)}$, and age, derived from best-fitting models.}

\begin{center}

                      \begin{tabular}{l@{\hskip 0.2in} @{\hskip 0.2in}c@{\hskip 0.2in}  @{\hskip 0.2in}c@{\hskip 0.2in}  @{\hskip 0.2in}c @{\hskip 0.2in}c}
                        \hline
                        \hline 
                        \rule{0pt}{3.4ex}
                        No-leakage & 12 + ${\rm \log(O/H)}$ & ${\rm \log(U)}$ & ${\rm \log(n_H/cm^3)}$, & ${\rm \log(Age/yr)}$ \\ 
                        \hline 
                        Single & 8.58 $\pm$ 0.20 & $ -1.92 \pm 0.69 $  & $ 2.91 \pm 0.47 $  &  $ 7.44 \pm 0.51 $  \rule{0pt}{3.4ex}\\
                        Binary & 8.60 $\pm$ 0.28 & $ -3.36 \pm 0.28 $ & $ 1.93 \pm 0.94 $  & $ 7.56 \pm 0.29 $ \rule{0pt}{3.4ex}\\
                       
                        \hline 
                        \hline
                        \multicolumn{2}{c}{ }  \\

\end{tabular}\\

\end{center}
\label{tab:parameters_ave}
%\end{table*}

%\begin{table*}

\caption{Including ionizing photon leakage, the average value and standard deviation of oxygen abundance 12 + ${\rm \log(O/H)}$, ionization parameter ${\rm \log(U)}$, hydrogen density ${\rm \log(n_H/cm^3)}$, and age, derived from best-fitting models.}

\begin{center}

\begin{tabular}{l@{\hskip 0.2in} @{\hskip 0.2in}c@{\hskip 0.2in}  @{\hskip 0.2in}c@{\hskip 0.2in}  @{\hskip 0.2in}c @{\hskip 0.2in}c}
                        
                        \hline
                        \hline 
                        \rule{0pt}{3.4ex}
                        Leakage & 12 + ${\rm \log(O/H)}$ & ${\rm \log(U)}$ & ${\rm \log(n_H/cm^3)}$, & ${\rm \log(Age/yr)}$ \\ 
                        \hline 
                        Single & 8.46 $\pm$ 0.32 & $ -3.27 \pm 0.40 $  & $ 2.48 \pm 0.53 $  &  $ 6.22 \pm 0.33 $  \rule{0pt}{3.4ex}\\
                        Binary & 8.75 $\pm$ 0.24 & $ -3.36 \pm 0.38 $ & $ 1.48 \pm 0.61 $  & $ 7.09 \pm 0.59 $ \rule{0pt}{3.4ex}\\
                        \hline
                        \hline
                        \multicolumn{2}{c}{ }  \\

\end{tabular}

\end{center}
\label{tab:parameters_ave_leak}
\end{table*}
\section{Best-fitting Models - no leakage}

\subsection{Oxygen abundance measurement}

Our best-fitting models are selected based on the emission line ratios and our resultant oxygen abundance for each model is  determined by the initial chemical composition for the model set as described in \cite{2018MNRAS.477..904X}. Here we can test the reliability of our best-fitting models by the comparison of the model abundance with those using O3N2 calibration as shown in Figure~\ref{fig:OH_O3N2}. In general, our binary-star models match the oxygen abundance in O3N2 calibration better, although there is still considerable scatter around the line of equality. Single-star models show less variance in the distribution and have their oxygen abundance  mostly around 8.5. This result may suggest that our final fit for single-star populations is likely to be insensitive to the metallicity compared to the results from our binary-star populations and their best-fits are more dependent on other parameters of the emission nebular that will be discussed in detail in the following section. A lower metallicity model is probably preferred to match the observed emission line ratios to allow for hotter stars in the stellar population that are present in the binary populations. Given the correlation of our model input metallicity with the oxygen abundance in O3N2 calibration as shown in \cite{2018MNRAS.477..904X}, this calibration is a relatively good estimation of oxygen abundance for nearby H\,II regions with moderate metallicity. In particular our models suggest significantly higher metallicity than derived from O3N2 calibrations for the type II SN hosts with lower metallicities. Therefore, our models predict similar oxygen abundance of the two SN types in a narrow range, which may indicate that the metallicity of the two SN types may not differ much in nearby galaxies.

\begin{figure*}
%\vspace*{-0.5cm}
\centering
\hspace*{-0.5cm}\includegraphics[width=8.5cm]{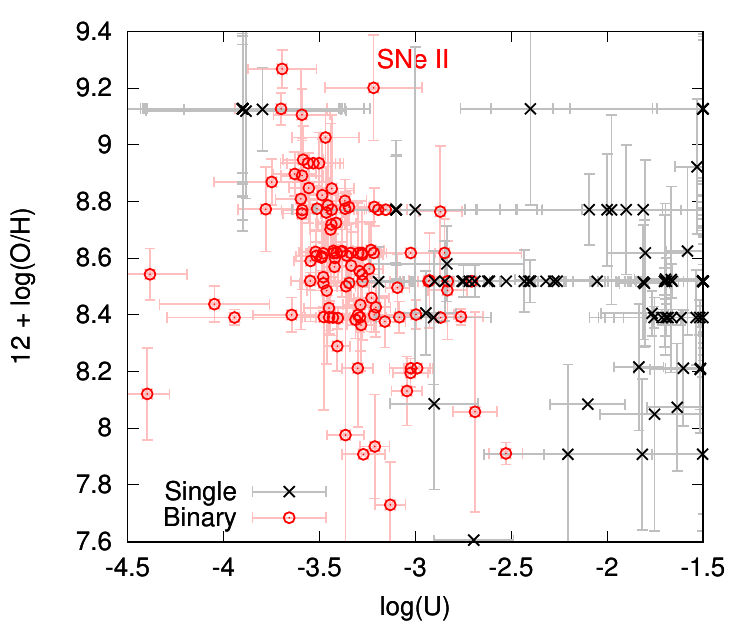} %\hspace*{0cm}
\includegraphics[width=8.5cm]{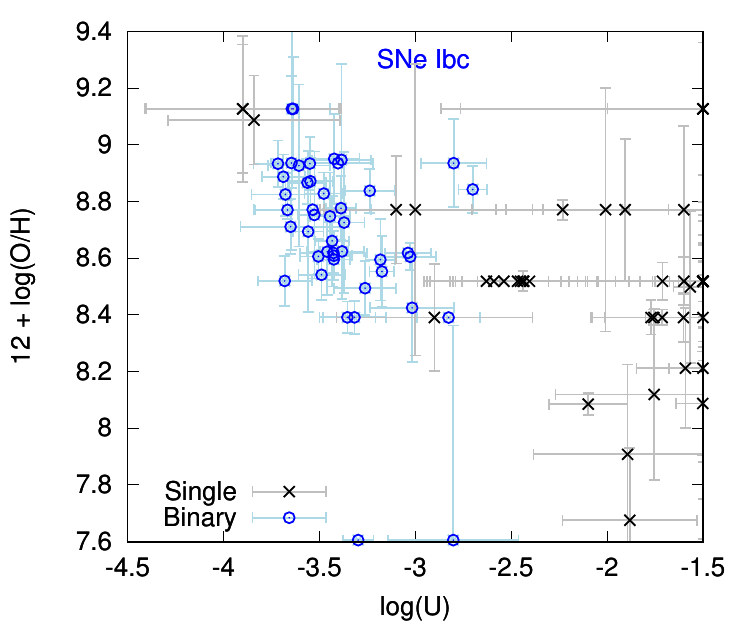} \\
\hspace*{-0.5cm}\includegraphics[width=8.5cm]{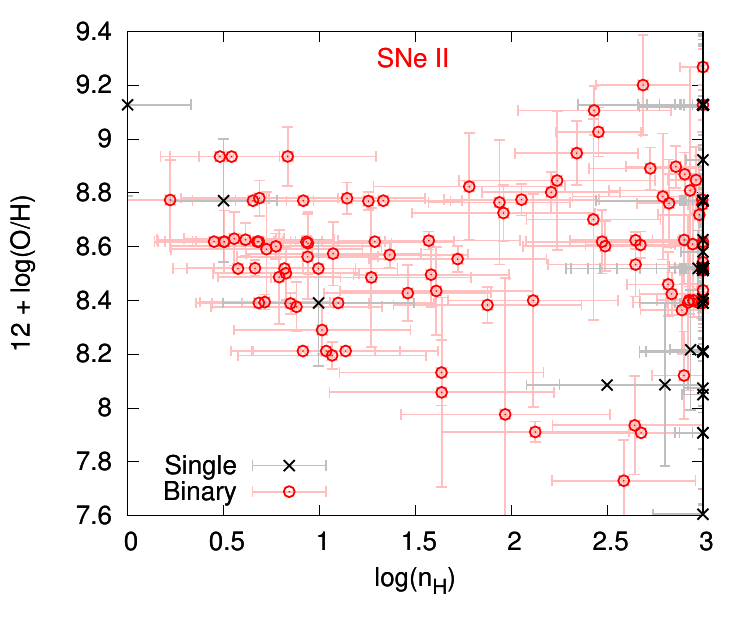} %\hspace*{0cm}
\includegraphics[width=8.5cm]{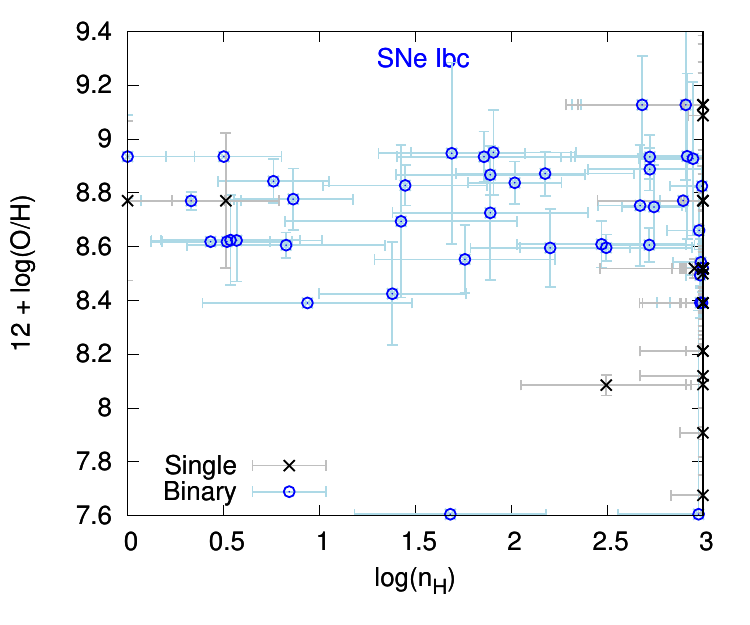} \\
\hspace*{-0.5cm}\includegraphics[width=8.5cm]{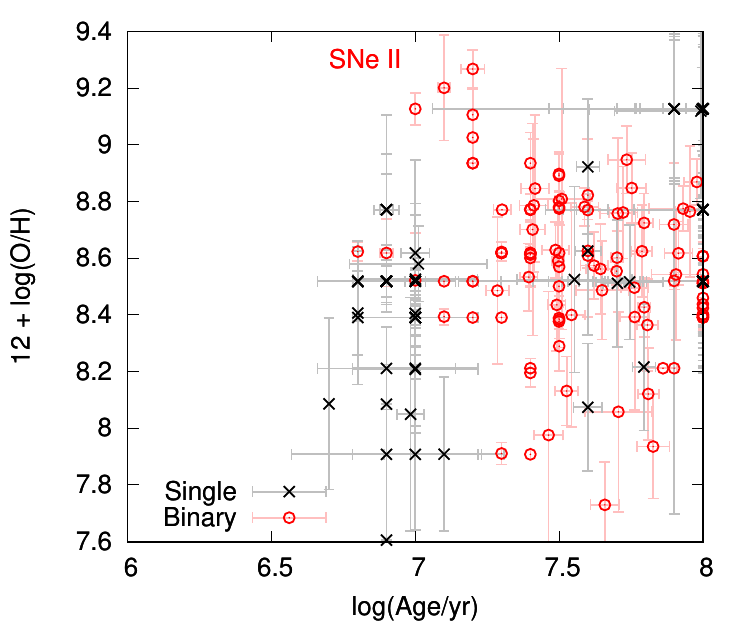} %\hspace*{0cm}
\includegraphics[width=8.5cm]{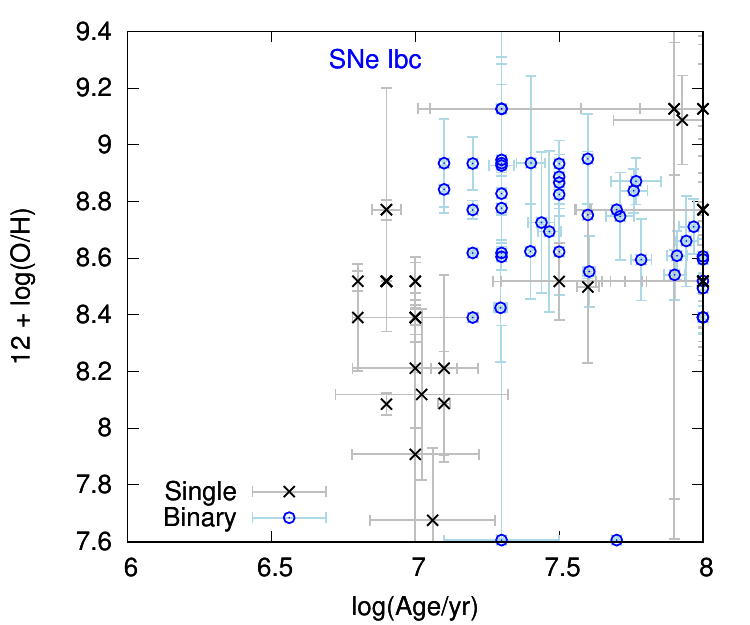} 
\centering
\caption{The variation of oxygen abundance with respect to ionization parameter ${\rm \log(U) }$ for the SN hosts in the top panels, ${\rm \log(n_H/cm^{3}) }$ in the middle panels and ${\rm \log(Age/yr) }$ in the bottom panels, with the left panels for type II hosts and the right panels for type Ibc hosts. Single-star models are in black crosses with error bars and binary-star models in red (SNe II) and blue crosses (SNe Ibc) with error bars.} \label{fig:UNA_SNe_1kpc}
\end{figure*}

\subsection{Oxygen abundance variation with respect to ${\rm U}$, $ {\rm n_H} $ and Age}
While the oxygen abundances derived from our best-fitting models are important, the other input model parameters, ionization parameter $ U $, gas density $ {\rm n_{H}/cm^{3}} $ and the age of the inner ionizing stellar population source tells us more about the surrounding environment of the SN and its stellar siblings. Table \ref{tab:parameters_ave} provide an overview of the average of best-fit parameters. We note the similar average oxygen abundance and similar age between single and binary populations, but a larger oxygen abundance variance appears in binary models as discussed before and larger age variance in single population. In addition, binary population are distinguished from the single populations with the lower ionization parameter and higher $ {\rm n_H}$.

Figure~\ref{fig:UNA_SNe_1kpc} displays how oxygen abundance in our fits are related to these parameters for both SN types separately. We note that these SN nebular regions have a different ${\rm \log(U) }$ distribution for our binary models compared to that of expected for normal young H\,II regions like those in van Zee sample we have studied previously \citep{2018MNRAS.477..904X}. First, there is only a weak trend  between oxygen abundance and ${\rm \log(U) }$ distribution with most binary-star models  concentrated between ${\rm \log(U) }$ = -3.5 and -3.0, with higher values at lower metallicities. This suggests a significantly lower ionization state in these CCSN nebula regions than in young H\,II regions where ${\rm \log(U) }$ up to -1.5 is more typical. These lower ${\rm \log(U)} $ values are consistent with the CCSN regions' position in BPT diagrams, traced on the bottom region of the H\,II region sequence. For single-star models, they all have surprisingly high values of ${\rm \log(U) }$ varying from -3 to -1. For both model sets there is little or no difference between the SN types. 

In addition, the oxygen abundance variation with respect to the gas density $ \log({\rm n_H/cm^3}) $ is also shown in Figure~\ref{fig:UNA_SNe_1kpc}. For both type II and Ibc SNe, binary-star models vary over the full range of gas density. In contrast, most single-star models are concentrated at the highest hydrogen density in our model of $ \log({\rm n_H/cm^3}) = $ 3.0. We find this difference in density is closely correlated with ages of the inner ionizing source. Most typical H\,II regions are young with ages below 10\,Myr, these CCSN regions from the binary model fits mostly have ages beyond 10\,Myr. This is consistent with their lower ${\rm \log(U) }$ as less ionizing photon production occurs at later ages. In comparison all the single-star model fits pile up at 10\,Myr.

While the metallicity is determined primarily from the abundance of the gas being ionized the other parameters are mostly determined by the ionization photon spectrum from the stellar population. The difference here between the single-star and binary models is because the interacting binaries allow hot stars to exist at ages beyond 10\,Myr where in the single star models none exist. Therefore to achieve a fit the single star models have to resort to high gas densities, and high ionization parameters to match the observed emission-line fluxes. The ages all then sit around 10\,Myrs because this is the time when the EW of the lines decreases to match the observed values. Since the hardness of the ionizing spectra also drops off, a very short lived phase in the single star models satisfies both ionizing spectrum and line strength constraints for the majority of the models. By contrast, the evolution is more gradual and there is a wider range of possible ages in the binary models. The values from the binary star fits appear to be more sensible and relatistic when compared to the estimated age of the stellar populations. Importantly we detect no SNe that occur in a very young H\,II region hosting very massive young stars. There is a strong consensus here between single and binary star models that different types of SNe all arise from progenitors with ages beyond 10\,Myr, equivalent to less-massive progenitors with initial masses less than 20\,M$_{\odot}$. 

\begin{figure*}
\centering
%\hspace*{-0.5cm}\includegraphics[width=8.3cm]{./figs_1kpc/OH_O3N2_sin_leakage_1kpc.pdf}
%\hspace*{-0cm}\includegraphics[width=8.3cm]{./figs_1kpc/OH_O3N2_bin_leakage_1kpc.pdf}\\
\textbf{Including ionizing photon leakage}\\
\hspace*{-0.5cm}\includegraphics[width=8.3cm]{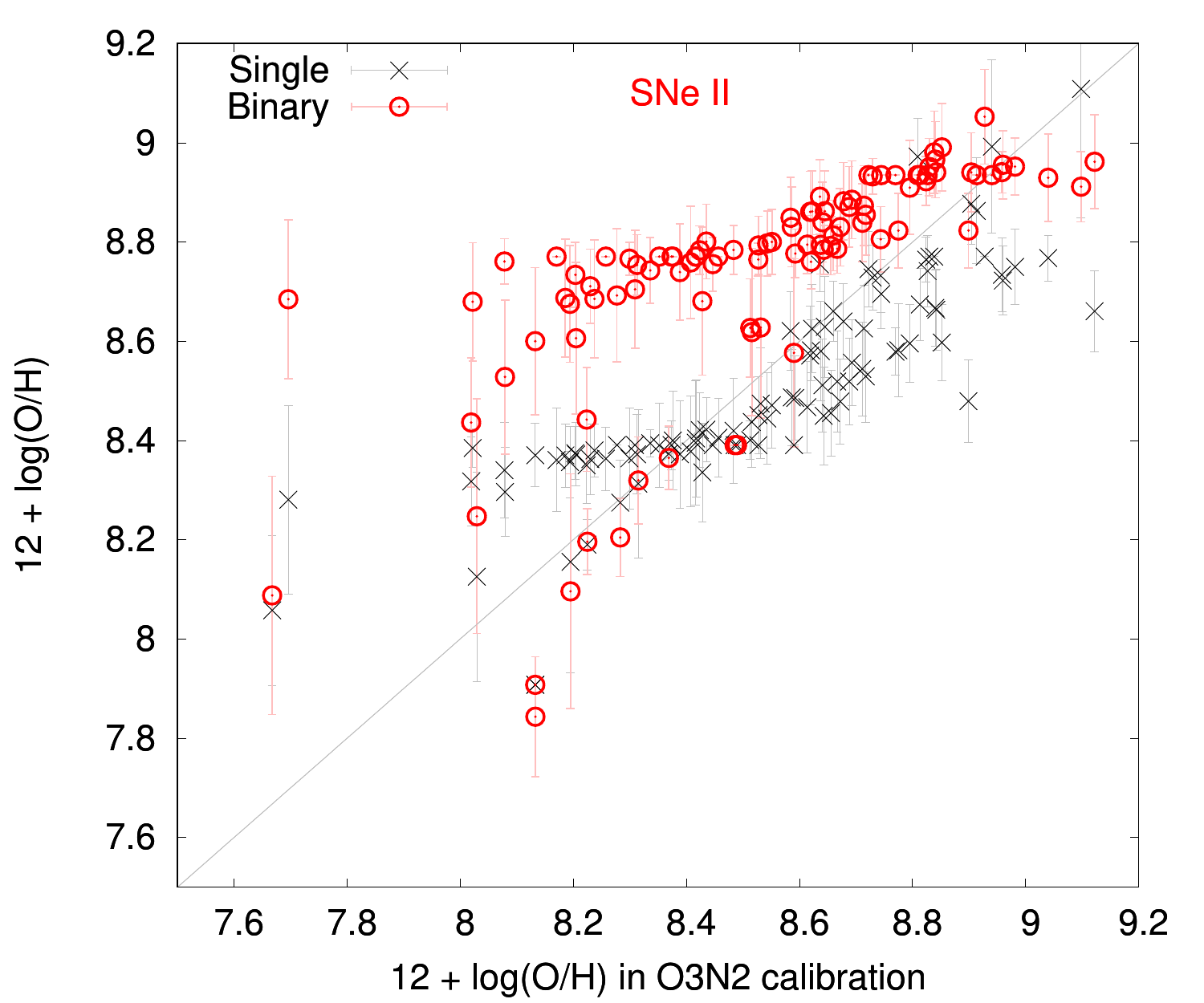}
\hspace*{-0cm}\includegraphics[width=8.3cm]{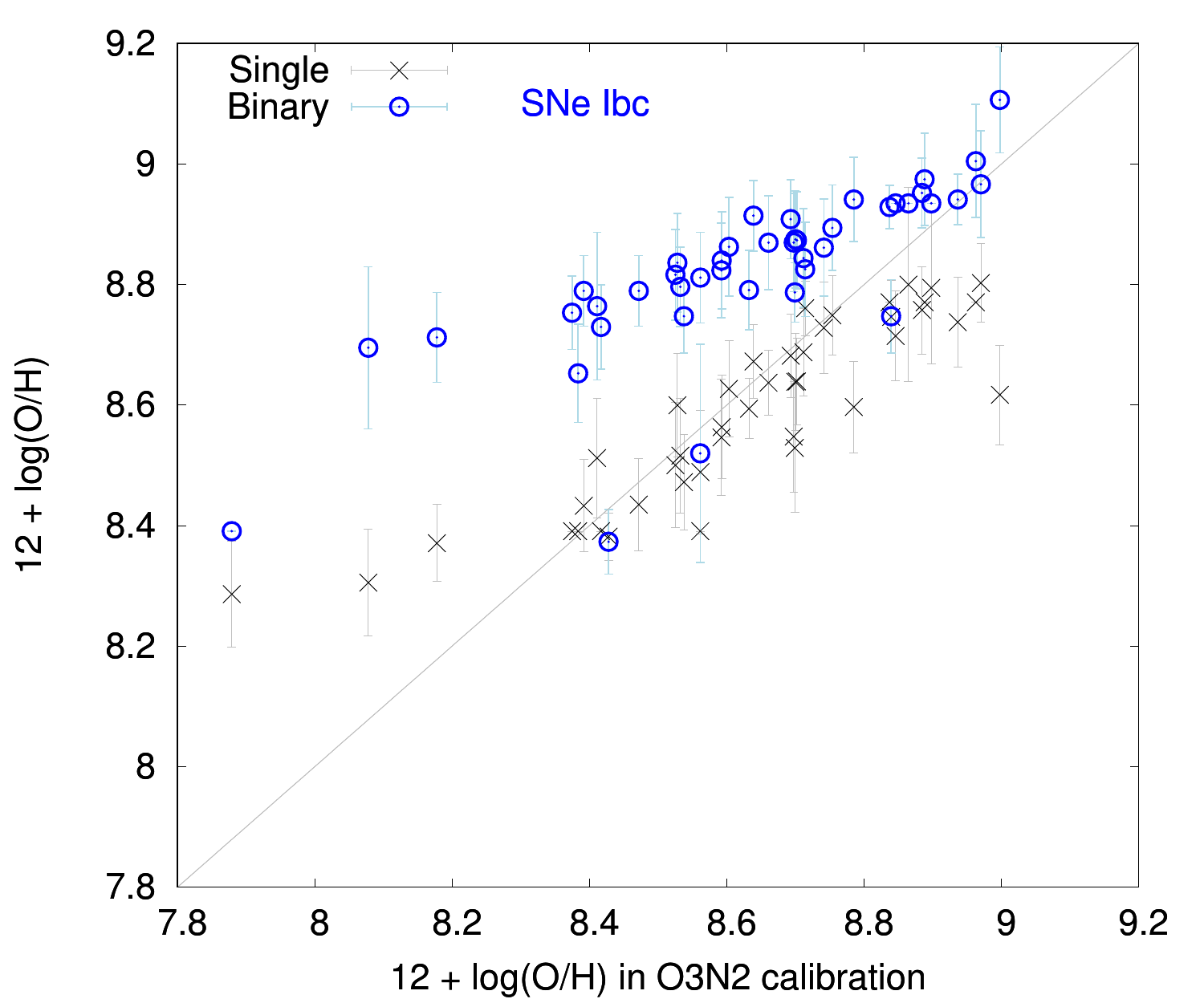}
\centering
\caption{In ionizing photon leakage case, predicted oxygen abundance from best-fitting models comparing with those using O3N2 calibration. It shows the comparison between single- and binary-star models separately for type II SN hosts in the left and type Ibc SN hosts in the right. Single-star models are presented by the black crosses with error bars and binary models are the red (SNe II) and blue (SNe Ibc) crosses with error bars.} \label{fig:OH_O3N2_leakage_sb}
\end{figure*}

\section{The Effect of Ionizing Photon Leakage}

In the above analysis we have assumed that every ionizing photon emitted from the stellar population contributes to the emission nebula flux. However we know that in many regions some of these ionizing photons are lost, either by being absorbed and heating dust grains, or simply escaping from the gas cloud before they have a chance to interact. In \cite{2018MNRAS.477..904X} we investigated the effect of ionizing photon leakage on reducing emission line strength and how this altered the fit of models to H\,II regions. In general we found that the best-fit age with leakage included leads to the generally younger H\,II regions, where the high EW values of the models meant they were otherwise disfavoured in the fitting calculation. In this section we allow for this leakage or loss of ionizing photons to see if the derived ages change.

\subsection{Oxygen Abundance}

First, including ionizing photon leakage has a strong affect on the oxygen abundance measurement as shown in Figure~\ref{fig:OH_O3N2_leakage_sb}. There are clear differences in oxygen abundance estimation compared to previous models without ionizing photon leakage. Both model sets now have a more linear relationship to the oxygen abundance estimated from the O3N2 method. However there is a significant off-set between the single-star and binary models. The difference can be explained in that to achieve the same strength of the ionizing spectrum single stars prefer lower metallicities where the average stellar temperatures are higher. The binary models on the other hand have hotter post-main sequence stars, even at higher metallicities, so do achieve a reasonable fit without needing to decrease to a low metallicity. As a consequence, allowing leakage makes the metallicity predictions from the O3N2 method consistent with single-star models, due to the fact that as varying metallicity is the primary way to vary the hardness of the ionizing radiation from single stars. For the binaries metallicity is a small effect since the helium stars produced at late ages the ionizing radiation hardness varies less. Nevertheless we again see no clear distinction between the range of host metallicity in the two CCSN types. Therefore, we can suggest the similar ionizing photon leakage effect for these two CCSN hosts is probably due to similar ISM distribution within their host H\,II regions. 

Compared to the average oxygen abundance and other physical parameters of the best-fitting models listed in Table \ref{tab:parameters_ave}, Table~\ref{tab:parameters_ave_leak} stresses again that including ionizing photon leakage can change these best-fitting parameters significantly. Binary-star models increase their average oxygen abundance by nearly 0.2 dex, while the average for single-star models decreases by about 0.1 dex. For single-star models, this change in oxygen abundance is combined with $ {\rm \log(U) } $ a factor of two lower and a significant decrease in age by 1.2 dex, while little difference in hydrogen density $ {\rm \log(n_H/cm^3) }$. The binary-star models also experience a corresponding decreases in age by 0.5 dex, but have almost no change in $ {\rm \log(U) } $ and a slight decrease in $ {\rm \log(n_H/cm^3) } $. Therefore, we suggest that the most important impact of ionizing photon leakage are the reduce of age and corresponding variance in oxygen abundance. 

\begin{figure*}
%\vspace*{-0.5cm}
\centering
\textbf{Including ionizing photon leakage}\\
\hspace*{-0.5cm}\includegraphics[width=8.5cm]{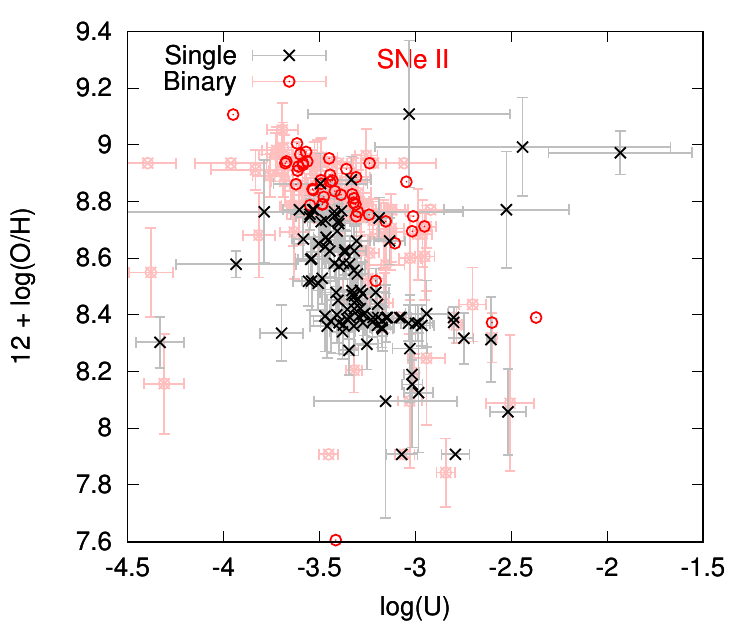}
\hspace*{-0cm}\includegraphics[width=8.5cm]{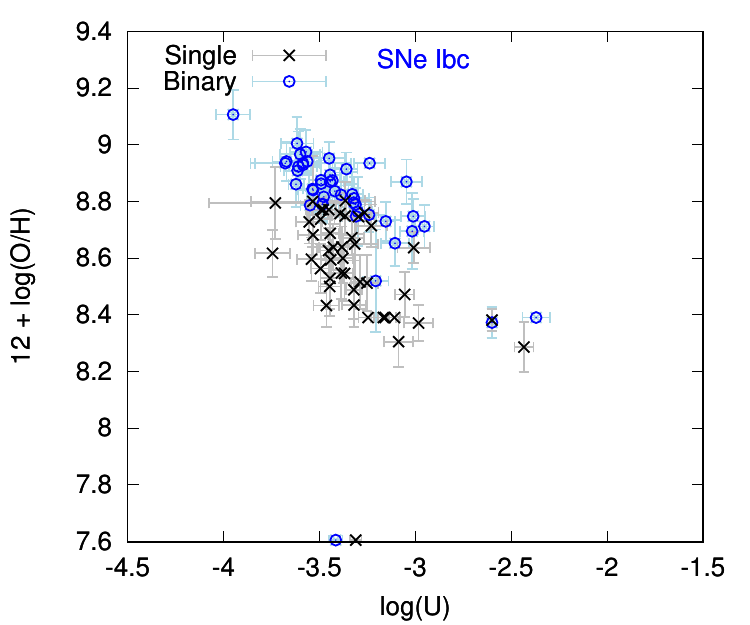}\\
\hspace*{-0.5cm}\includegraphics[width=8.5cm]{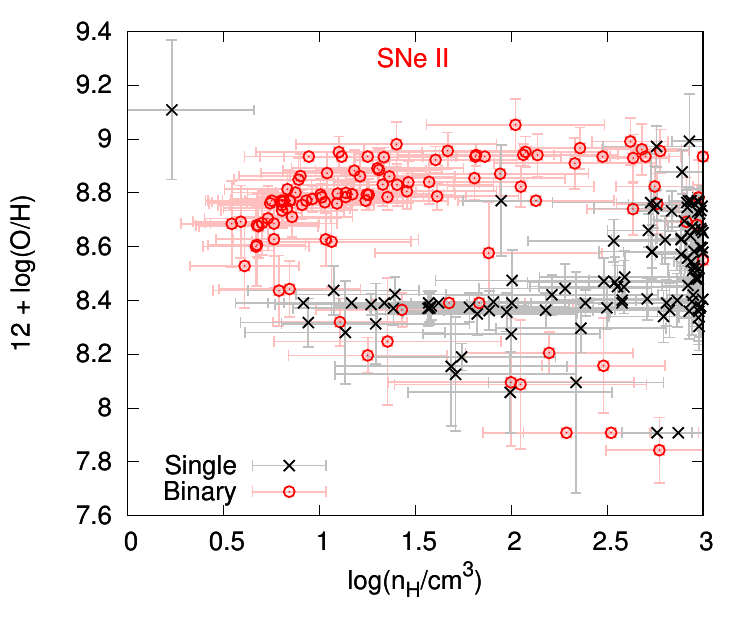}
\hspace*{-0cm}\includegraphics[width=8.5cm]{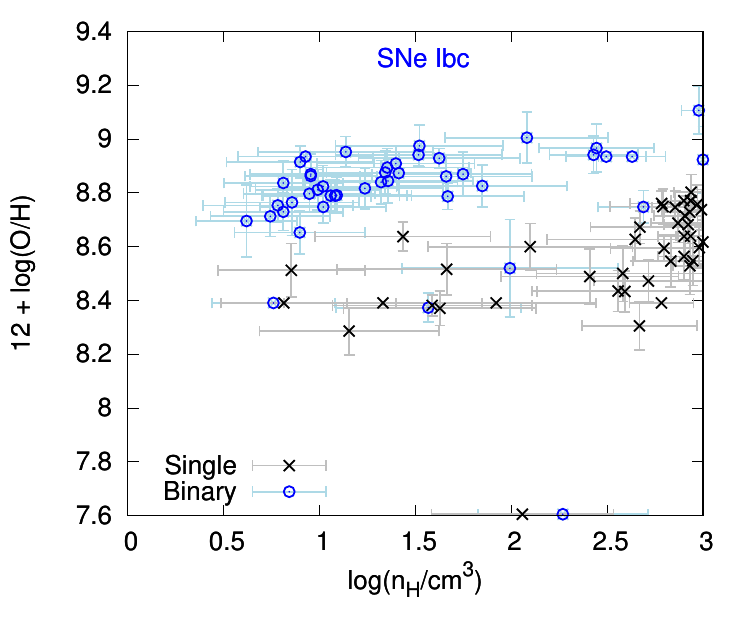}\\
\hspace*{-0.5cm}\includegraphics[width=8.5cm]{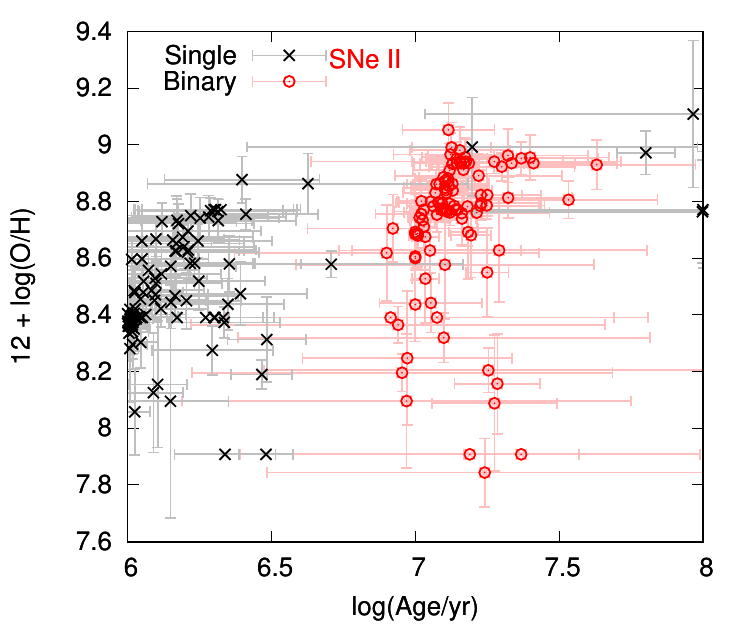}
\hspace*{-0cm}\includegraphics[width=8.5cm]{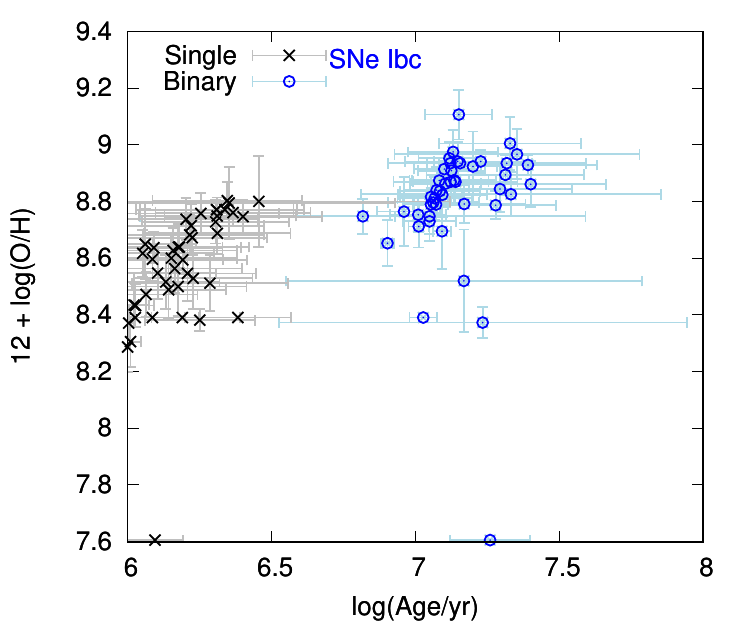}
\centering
\caption{In ionizing photon leakage case, the variation of oxygen abundance with respect to ionization parameter $ {\rm \log(U) }$ for the SN hosts in the top panels, ${\rm \log(n_H/cm^{3}) }$ in the middle panels and ${\rm \log(Age/yr) }$ in the bottom panels, with the left panels for type II hosts and the right panels for type Ibc hosts. Single-star models are in black crosses with error bars and binary-star models in red (SNe II) and blue circles (SNe Ibc) with error bars.} \label{fig:UNA_leakage}
\end{figure*}

\begin{figure}
%\vspace*{-0.5cm}
\centering
\hspace*{-0.5cm}\includegraphics[width=9cm]{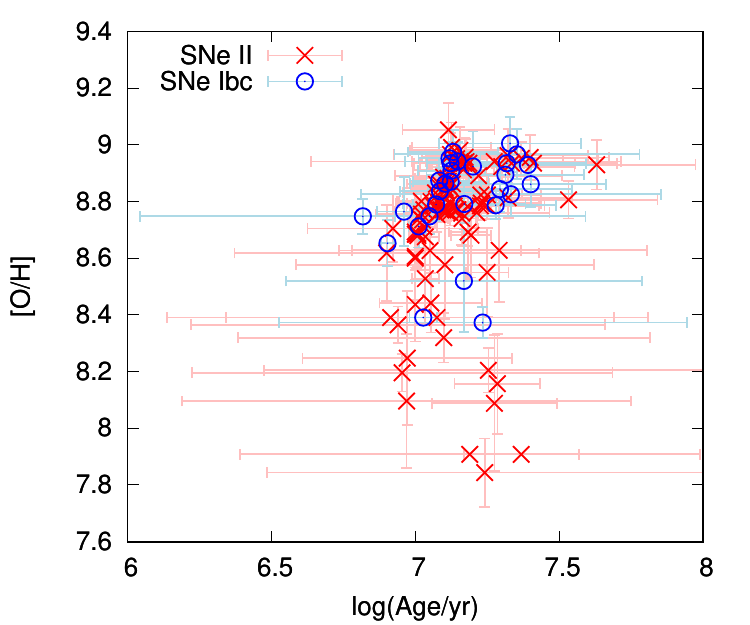}\\
\centering
\caption{Age constraints on type II and type Ibc SN progenitors from binary-star models. Red crosses with error bars are for type II SNe and blue circles with error bars are for type Ibc SNe.} \label{fig:Age_SNe_bin}
\end{figure}

\subsection{Oxygen abundance variation with respect to ${\rm U}$, $ {\rm n_H} $, and age}

When we consider the other model parameters we obtain from our fit we again find that including ionizing-photon leakage leads to very different results. Again the distributions of these two CCSN types are very similar but show differences compared to the previous result shown in Figure \ref{fig:UNA_leakage}. Now single-star models have shifted to lower ionization parameters, with ${\rm \log(U) }$ between -4 and -3, just as for the binary-star models. In addition, we find the best-fitting models including ionizing photon leakage show more preference for lower gas densities compared to previous models, although a significant fraction of the binary star models still prefer a high density with $ \log({\rm n_H/cm^3}) =3$. The most significant difference is that the single-star models become much younger with ages all below 3\,Myr except a few outliers. This also occurs for the binary-star models but more moderately so; most of the ages are between 10 to 30\,Myr. All the single-star models are too young to be a reasonable age estimate of the SN progenitors. Even the most massive stars have a minimum evolution lifetime of the order of 3\,Myr. 

The binary-star model age and metallicity range as shown more clearly in Figure~\ref{fig:Age_SNe_bin}. These agree with similar observational constraints for detected SN progenitor detections \citep{2015PASA...32...16S} and from resolved stellar populations \citep{2018arXiv180308112W}. These ages imply initial masses for most of the supernova in our sample of approximately 8 $ - $ 20 ${\rm M_{\odot} }$. This is approximately the mass above which it is expected that core-collapse produces black-holes that may not lead to visible supernova \citep[e.g.][]{2003ApJ...591..288H,2004MNRAS.353...87E,2017A&A...601A..29Z}. As $ {\rm H\alpha }$ EW is a sensitive tracer of age at higher metallicties (Z$>$0.004) as shown in Figure \ref{fig:EW_Ha}, the similar age distributions of the two SN types are incorperated with their almost identical distributions of $ {\rm H\alpha }$ EW as shown in Figure~\ref{fig:CD_EW}. In addition, we note that while the bulk of SNe of both types are seen across a similar range of metallicities, the fraction of events does appear to change with metallicity. At metallicities below 12 + log(O/H) $ \sim $ 8.5, the number of type Ibc events drops off sharply, while type II SNe drop off less rapidly and so become relatively more common. 
%We note that there is also a strong metallicity preference for type Ib/c to arise from metallicities close to solar with most low metallicity events being of type II. 

\subsection{The extent of remaining ionizing photons}
\begin{figure}
\centering
%\vspace*{-0.5cm}
\vspace*{-0.5cm}
\hspace*{-0.5cm}\includegraphics[width=9.5cm]{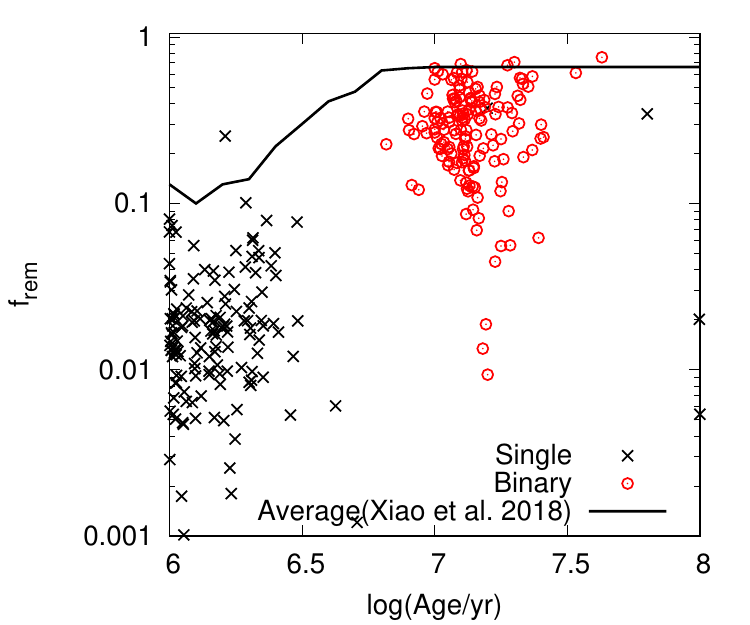}
\centering 
\caption{The fraction of remained ionizing photon leakage, $f_{rem}$, as a function of age. The black crosses are for single-star models and the red circles are for binary-star models. The solid line indicates the fractions inferred using an identical analysis for a large sample of typical H\,II regions in spiral and dwarf galaxies by Xiao et al. (2018) for comparison.}\label{frem_Age}
\end{figure}

\begin{figure}
\centering
%\vspace*{-0.5cm}
\vspace*{-0.5cm}
\hspace*{-0.5cm}\includegraphics[width=9.5cm]{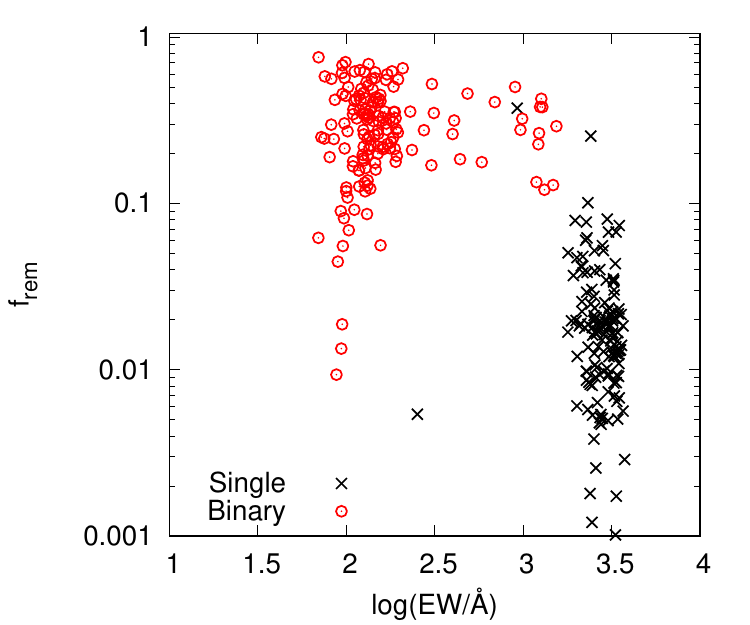}
\centering 
\caption{The fraction of remained ionizing photon leakage, $f_{rem}$, as a function of EW. The black crosses are for single-star models and the red circles are for binary-star models.}\label{frem_EW}
\end{figure}

Given that a large fraction of ionizing photons must escape to provide the best fits to the data, it is interesting to consider the fraction of remaining ionizing photons, which are absorbed and processed by the nebular gas. \cite{2018MNRAS.477..904X} discussed the remaining ionizing photon fraction as a function of age in HII regions in nearby dwarf and spiral galaxies. Here, figures \ref{frem_Age} and \ref{frem_EW} present the fraction of photons remaining to interact with the nebular gas after accounting for photon leakage, $f_{rem}$, as a function of age and EW respectively for CCSN host regions. Compared to the equivalent fraction in \cite{2018MNRAS.477..904X}, the CCSN hosts experience more ionizing photon leakage, with roughly an order of magnitude lower $f_{rem}$ for single-star models, while in the binary model fits, the typical remaining fractions are a factor of two lower than in our previous study. This is a corollary of the young ages required in the single model fits - the observed data require a very high photon production rate to explain the ionization state but must then lose most of the photons in order to explain the recombination line equivalent width. For binary stars, already a more efficient source of hard ionizing photons, a smaller escape fraction is required. The 50 per cent ionizing photon leakage typically required by binary-star models is consistent with the fact that CCSN hosts are relatively old and more defused nebulae, through which ionizing photons may have a sufficiently long mean free path to escape without interaction.

\section{Discussion}
\begin{figure}
\centering
%\vspace*{-0.5cm}
\vspace*{-0.5cm}
\hspace*{-0.5cm}\includegraphics[width=9.5cm]{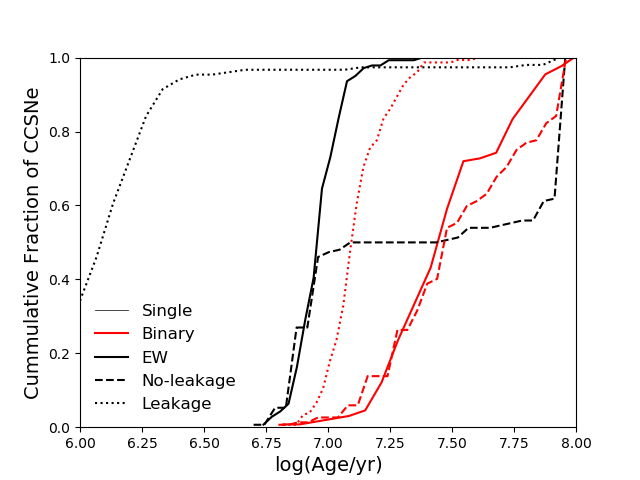} %frac_sb_1kpc-BW.pdf
\centering 
\caption{Cumulative fraction of CCSNe as a function of age derived in three methods: directly from H$ \alpha $ EW (solid lines), best-fitting models without ionizing photon leakage (dashed lines) and best-fitting models including leakage (dotted lines). The single-star models are in the black lines and binary-star models are in red lines. }\label{frac_sb_SN2}
\end{figure}

\begin{figure}
\centering
%\vspace*{-0.5cm}
\vspace*{-0.5cm}
\hspace*{-0.5cm}\includegraphics[width=9.5cm]{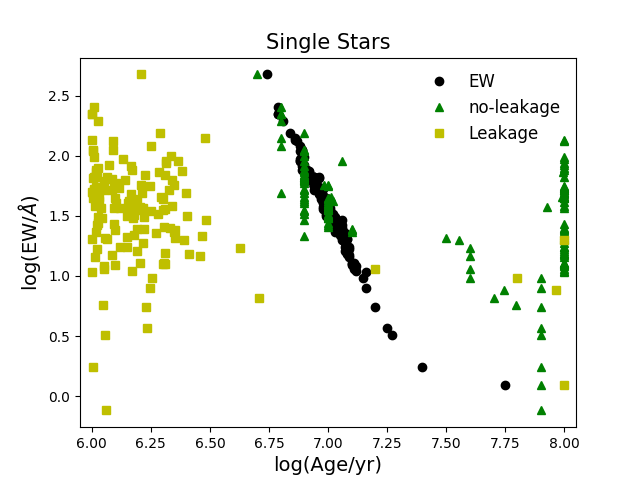}
\hspace*{-0.5cm}\includegraphics[width=9.5cm]{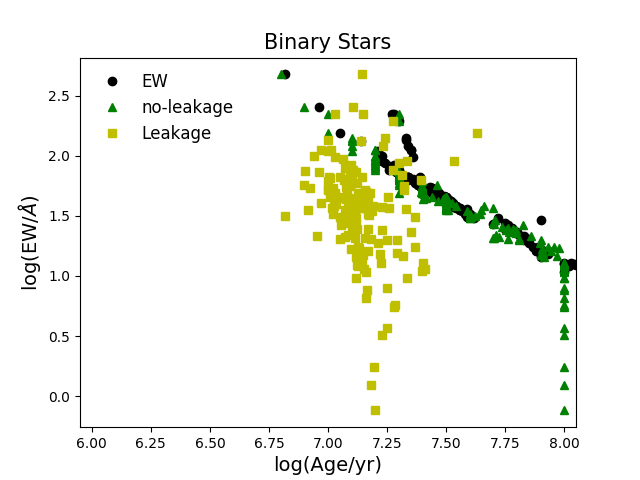}
\centering 
\caption{The EW of H${\rm \alpha }$ vs derived age from EW-based method and best-fitting models with and without ionizing photon leakage. The black dots represent the EW-based ages, the green triangles and yellow squares are ages from no-leakage and leakage best-fitting models respectively.}\label{EWAge_sb}
\end{figure}

\begin{table*}
\caption{The comparison of ages derived by our method to those for SN were progenitor detections or constraints \citep{2015PASA...32...16S,2013MNRAS.436..774E} and surrounding stellar population constraints \citep{2016MNRAS.456.3175M, 2017MNRAS.469.2202M,2018MNRAS.476.2629M}.}
\begin{center}
\begin{tabular}{cccccccccc}
\hline
\hline
          & & Progenitor  & Progenitor & Maund stellar& No leakage & No leakage & Leakage & Leakage \\
          & & best estimate   & Lower limit  & populations &  Single  & Binary& Single  & Binary \\
SN        & Type& $\log({\rm age/yrs})$    & $\log({\rm age/yrs})$    & $\log({\rm age/yrs})$& $\log({\rm age/yrs})$    &  $\log({\rm age/yrs})$ &$\log({\rm age/yrs})$     &  $\log({\rm age/yrs})$\\
\hline
\hline
2003gd  &IIP &7.7 &  $>$7.3  & 7.14, 8.08 & 6.9$\pm$0.0 & 7.49$\pm$0.05  & 6.30$\pm$0.19  &  7.08$\pm$0.73   \\
2013ej  &IIP &7.4 &  $>$7.2  & 7.16, 8.03 & 8.0$\pm$0.38 & 7.93$\pm$0.12 & 6.12$\pm$0.26  &  7.13$\pm$0.07  \\
1999gi  &IIP &--   & $>$7.2  & -- & 6.9$\pm$0.04 & 7.4$\pm$0.0  & 6.40$\pm$0.27  &  7.18$\pm$0.54   \\
1999em  &IIP &--   & $>$7.1  & -- & 8.0$\pm$0.78 & 8.0$\pm$0.0 & 6.10$\pm$0.24  &  7.12$\pm$0.02  \\
\hline
2007gr  &Ic &6.8 or 7.3 & $>$6.8 & 6.79, 7.69 & 8.0$\pm$0.7 & 7.94$\pm$0.04 & 6.19$\pm$0.22 & 7.17$\pm$0.21 \\ % Mazzali et al. (2010) suggest an initial mass of 15Msun or an age of 7.1 mention in text.
2001B   &Ib  &--  & --   & 6.55, 7.14  & 7.0$\pm$0.22 & 7.3$\pm$0.04 & 6.21$\pm$0.24 & 7.09$\pm$0.15  \\
\hline
\hline
\end{tabular}
\end{center}
\label{tab:progenitorcomparison}
\end{table*}

In this paper we have attempted to constrain the age and metallicity of nearby SN progenitors by modelling the nebular emission lines from the stellar populations surrounding the SN site. Using single-star stellar populations, We have found that emission line ratios imply a combination of unrealistically dense gas, young stellar populations and high ionization parameters. In comparison models that take account of interacting binary stars provide a much better and more realistic parameters for the SN progenitors. The estimated model parameters have lower gas densities, lower ionization parameters and older ages that would be expected for stellar populations where the bulk of SNe should occur. The nebular regions hosting CCSNe must already have lost their most massive stars, which collapse to black holes and are unlikely to form optically-luminous transients, or alternatively evolve into ultra-luminous and very rare transients. Instead they must be old enough to reach the terminal age of somewhat less massive stars which die as normal supernovae. They should thus differ from the very young, often ultraviolet- or H${\rm \alpha}$-selected, star-forming regions which are dominated by the most massive stars and are more conventionally identified by H\,II region surveys. These regions have high ionization parameters and gas densities \citep[e.g][]{2018MNRAS.477..904X} and so the fact that our binary model fits are not like this is consistent with what we should expect. 

\begin{figure*}
\centering
%\vspace*{-0.5cm}
\vspace*{0.5cm}
\text{H$ \alpha $ EW as age indicator}\\
\includegraphics[width=8.5cm]{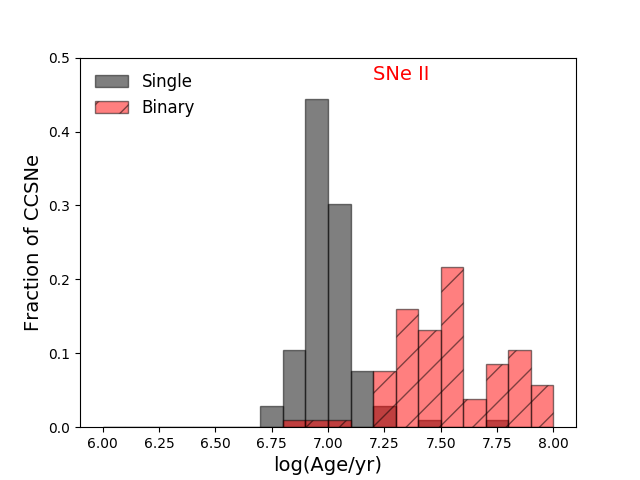}
\includegraphics[width=8.5cm]{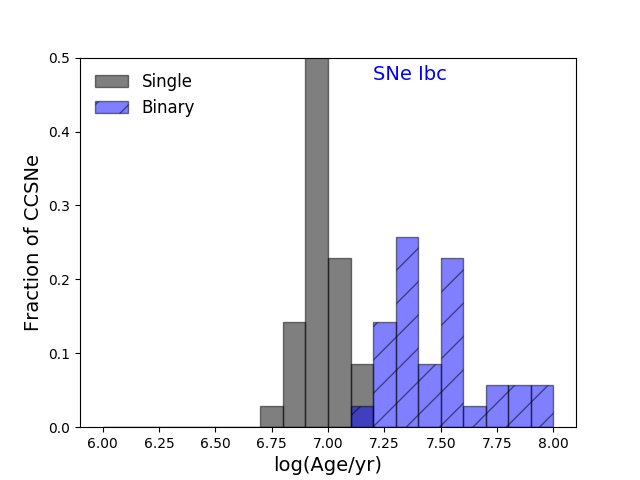}\\
\text{Age estimation in no-leakage case}\\
\includegraphics[width=8.5cm]{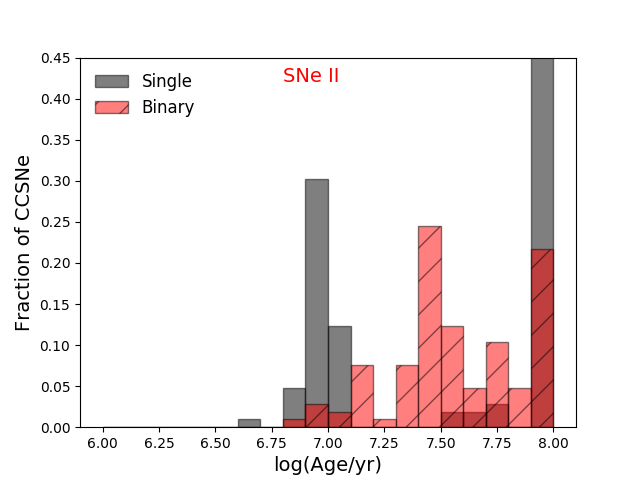}
\includegraphics[width=8.5cm]{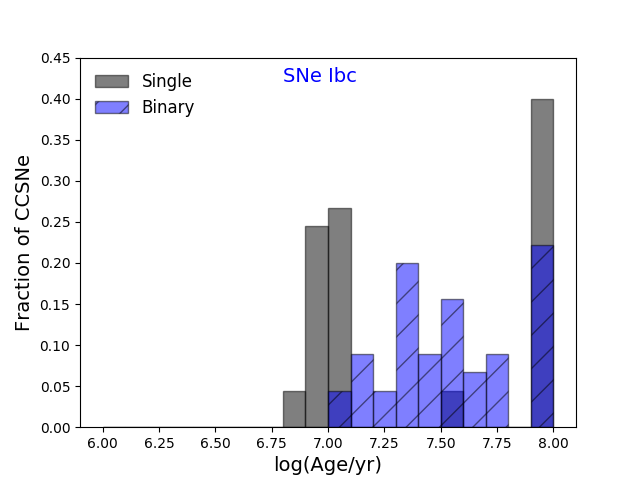}\\
\text{Age estimation in leakage case}\\
\includegraphics[width=8.5cm]{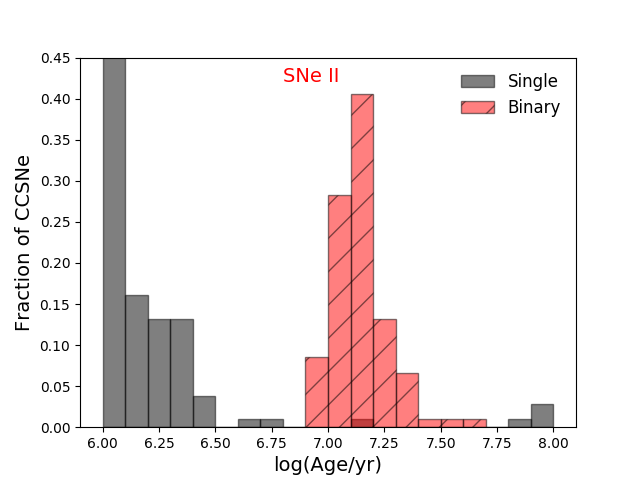}
\includegraphics[width=8.5cm]{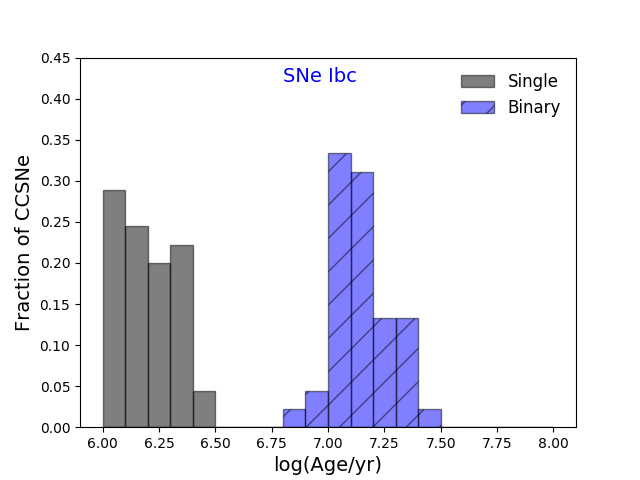}
\centering 
\caption{The fraction of CCSNe as a function of age derived in three methods: directly from H$ \alpha $ EW (top panels), best-fitting models without ionizing photon leakage (middle panels) and best-fitting models including leakage (bottom panels). The left panels show the fraction distribution for SNe II and the right panels for SNe Ibc. The black bars are for single-star models and red (SNe II) and blue  pattern (SNe Ibc) bars are for binary-star models. }\label{hist_SN1}
\end{figure*}

An important question is to decide whether we should favour our derived parameters assuming leakage or no leakage. To answer this question we first consider Figures~\ref{frac_sb_SN2}, \ref{EWAge_sb} and \ref{hist_SN1} that compare the predicted age of the SNe from the three methods to determine the age, H$\alpha$ EW and emission-line fitting without and with ionizing photon leakage. We see that the H$\alpha$ EW age and no-leakage emission-line fits agree closely. This reflects the fact that Ha EW drives the age in our fitting algorithm, while the line ratios are more sensitive to alternate parameters. However allowing for leakage decreases the ages derived from the emission-line fits. For the single star models these ages are unrealistically young at 3\,Myr and below. While for the binary models the change does allow the ages to be more plausible, with most lying between 10 and 30\,Myr. 

Theoretical predictions on the age distribution of SNe in both \citet{2017A&A...601A..29Z} and \citet{2017PASA...34...58E} show that the SN rate in binary-star models is highest at younger ages and decreases as older ages. This is because even though there are more lower mass stars, their evolutionary timescale becomes more dependent on mass and so they are spread out over a larger time. We see that for our preferred binary models assuming leakage there are some SNe with ages down to 6\,Myr the majority have ages of 10\,Myr and above. This agrees with the initial mass range at which there is a change in remnant formation from neutron stars to black holes, at round 20\,M$_{\odot}$ \citep{2004MNRAS.353...87E}, although we note that the situation might not be a simple as there being one single cut-off mass \citep[e.g.][]{2014ApJ...783...10S,2016ApJ...818..124E}. In the no-leakage case the majority of our derived SN ages are beyond 30\,Myr and up to 100\,Myr when these should be the rarest events. Therefore this suggests that our derived ages with leakage are a more accurate age estimate for the SN progenitor. This leakage fraction will also incorporate a factor of dilution of the emission lines from any underlying old stellar population that may also be contributing the the continuum flux. Nevertheless, recent studies of SNe and supernova remnants (SNRs) both suggest the SN progenitors would tend to be from older propolations. For example, \cite{2017A&A...601A..29Z} predict that about 15 per cent of CCSNe orgrinating from binary-star systems, are late to occur 50$-$200 Myr after birth. Then recently, \cite{2018ApJ...861...92D} analysing the stellar population surrounding 94 supernova remnants in M31 and M33, where they found some SNRs to have associated stellar populations older than 30 Myr, up to 80 Myr. Thus our findings add to the growing evidence that some core-collapse supernovae arise from older stellar populations.

An important deduction we can make from Figures~\ref{frac_sb_SN2} and \ref{hist_SN1} is why using single-star H$\alpha$ EW provides a reasonable age estimate for SN progenitors. The difference in the single-star EW ages and the binary models with leakage is only approximately 0.2 dex. Thus one simple conclusion we could make is that all SN progenitor ages derived by H$\alpha$ underestimate the stellar population ages by approximately 60 per cent and thus overestimate the initial progenitor mass \citep[e.g.][]{2017ApJ...849L...4C}. In Figure \ref{EWAge_sb} we show that it is possible to use our work to calibrate a crude but broadly correct H$\alpha$ EW to age relation. However this relation shows a large scatter due to the range of metallicities and gas conditions in the sample and we highly recommend that an age should be derived using a full set of nebular emission lines to provide greater accuracy.

It is also worthwhile to compare our age ranges to other estimates from progenitor and stellar population studies that are extant in the literature. We show the results of such a test in Table \ref{tab:progenitorcomparison}. Here we include the SNe where pre-explosion images exist that enable constraints on the initial masses of the progenitor stars to be placed. We have used the masses from \citet{2013MNRAS.436..774E} and \citet{2015PASA...32...16S}, converting the masses to estimated progenitor lifetimes using our stellar models. We also list the lower age limit from the progenitor detection. We also include age estimates by \citet{2016MNRAS.456.3175M, 2017MNRAS.469.2202M,2018MNRAS.476.2629M} who have used similar imaging to study the stellar population surrounding the SN sites. From this they derive the ages of the largest stellar populations which we list here. Although we note the ages were derived using single-star evolution tracks and ignore the effect of interacting binaries on the stellar population. Therefore these ages are likely to be underestimates \citep{2017PASA...34...58E}.

We see that most of the progenitor ages lie in the region of around a $\log({\rm age/yr})$ of 6.8 to 7.7. While the uncertainties are large there is consistency between these ages and those from our binary population ages. In most cases our single star estimates give values that are either too old or too young. It is however difficult to decide between whether we should include our leakage model or not. In general the binary no leakage model tends to provide a fit with low uncertainty that can be significantly old when the SN rate is expected to be low. With leakage the ages are younger, although with a higher uncertainty in some cases. We find this is still suggestive that our inclusion of allowing for leakage provides a more reasonable fit. We note there is other secondary evidence for SN 2007gr. \citet{2010MNRAS.408...87M} studied the light curve of 2007gr and found that the SN is consistent with the explosion of a 15${\rm M_{\odot}}$ star, this is consistent with a lifetime of $\log({\rm age/yr})=7.1$ which is consistent with our ages but not that of \citet{2016MNRAS.456.3175M}. The reason for this mismatch is likely to be due to the fact their age derivation was made using single-star evolution tracks. If a fit was carried out with a binary population \citep[e.g. as for Cygnus OB in][]{2017PASA...34...58E} then the age will likely become consistent with the older age.

A final view on our results is the relationship between, metallicity and progenitor age. We see in Figure \ref{fig:Age_SNe_bin} than in general the two different types of SNe cover a similar range of ages and oxygen abundance values but the oldest progenitors are of type II as well as the lowest metallicity events. This is in agreement with such predictions as \citet{2003ApJ...591..288H} and \citet{2004MNRAS.353...87E} because these stars experience less mass loss over their lives. The older, less massive stars have weaker winds while at lower metallicities line-driven winds are weaker for stars of the same metallicity.

\begin{figure*}
\centering
%\vspace*{-0.5cm}
\vspace*{-0.5cm}
\hspace*{-0.5cm}\includegraphics[width=18cm]{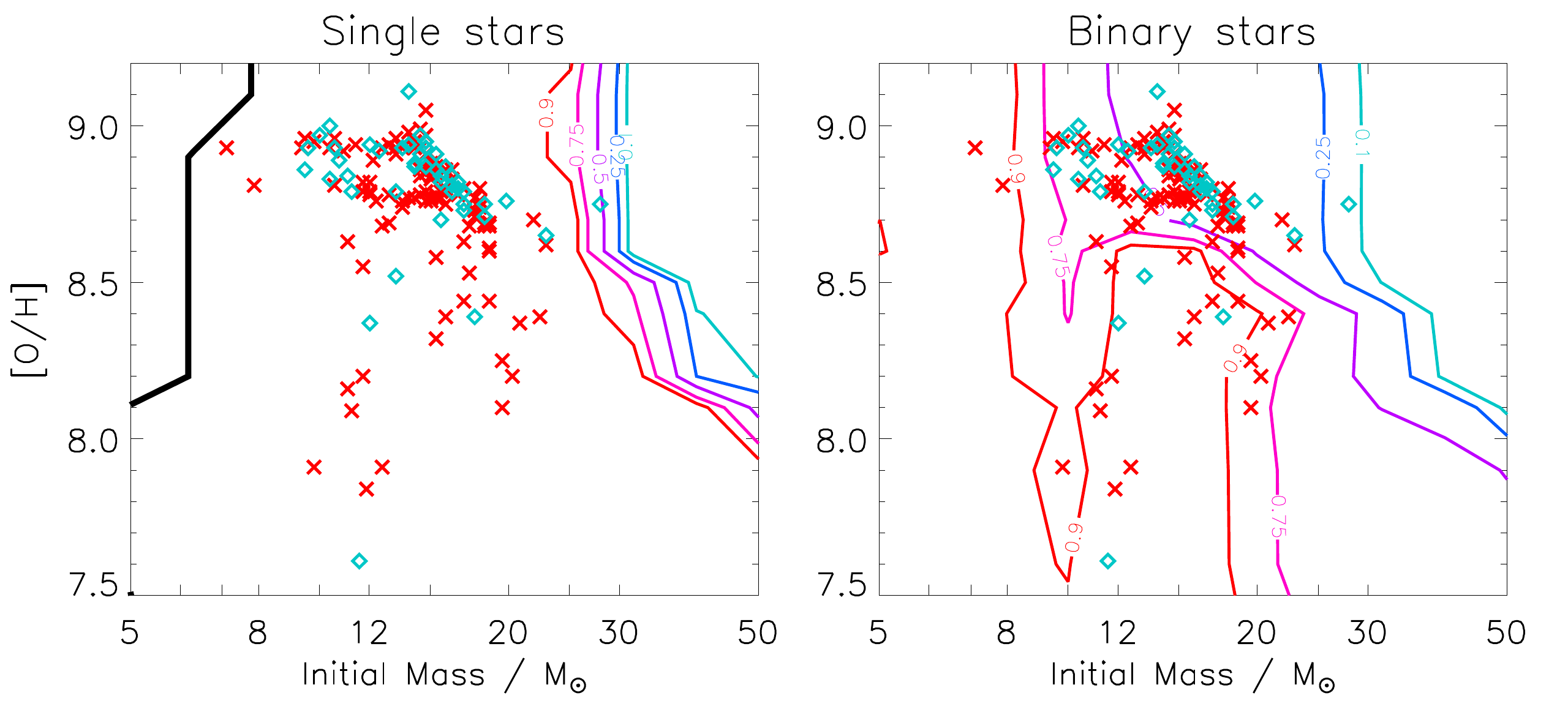}
\centering 
\caption{The predicted number count ratio of SN type as a function of initial mass and metallicity, compared to observational data (type II in red crosses, type Ibc in blue diamonds) from the best-fitting models accounting for ionizing photon loss. The contours are annotated with the fraction of supernovae that are type II at that contour. The thick black line represents the minimum initial mass for SNe to occur.}\label{hegerplot}
\end{figure*}

Another way to see how our derived ages and metallicities from the emission lines compare to model SN predictions is to convert our ages into effective initial masses. We use our single-star models to determine a relation between the age of a stellar population and the mass of a single-star with the same lifetime. Using these values we then created Figure \ref{hegerplot}, which is similar to the widely used figure in \citet{2003ApJ...591..288H}, which shows the regions where different SN types occur in initial mass versus initial metallicity space. These figures show similar structures to our earlier attempts at these figures in \citet{2004MNRAS.353...87E}. Here however we use oxygen abundance rather than metallicity mass fraction to measure the metal enrichment. The first panel which shows where different SN types occur for single stars demonstrates that for the initial masses and metallicities we derive all SNe in our sample should be type II. Therefore most SN progenitors cannot arise from single star progenitors.

Creating a binary version of these progenitor figures has the problem that there is no longer a direct connection between initial mass and final evolutionary outcome. This is because of the many diverse possible evolutionary pathways that become possible due to binary interactions. Our attempt at such a figure, shown in the second panel of Figure \ref{hegerplot}. It was created by looking at all stars from our binary models that have the same initial mass and initial metallicity. We use logarithmic mass bins of 0.1 dex width and use metallicity bins for our source models. Then for each model that lies in a specific bin that experiences a core-collapse SNe (if the final mass is $>$1.5 M$_{\odot}$, the CO core mass is $>$1.38 M$_{\odot}$ and core carbon burning has occured) we identify if they are type II or type Ib/c supernovae by the hydrogen mass in the model, if it is $>10^{-3}$ M$_{\odot}$ then the SN is of type II. From this at each point in the progenitor figure we can work out the fraction of SNe that are type II, with the remainder being type Ib/c. We then draw contours of these values of the initial parameter space, the resulting figure showing for which initial parameters type II SNe and type Ib/c SNe are most common. We see that at high masses and metallicities most of the SNe are type Ib/c due to stellar wind mass loss. At lower masses, especially at lower metallicities, most SNe are type II because even though binary interactions might remove a significant amount of surface hydrogen the stellar winds are unable to remove the remaining hydrogen before core collapse.

The location of the SN progenitor masses and metallicities we derive here broadly agrees with these models. A mix of events occur close to the even mix of SN types while more type II occur relative to type Ib/c where this is predicted. It is important to note that the same models that went into the creation of the contours in this figure are also those that are used to evaluate and characterise the  stellar population at the location of the SNe, and hence estimate the progenitors. However different aspects of those models were used. For example, how much hydrogen remains at the end of the progenitors' lives were used to estimate their type, while for the emission-line fitting we used the total luminous output of the stellar models at the same age. Therefore a key result from this figure is that the average SN types we infer from the surrounding emission line output of a living stellar population is related to the most likely SN type, i.e. if there are a number of stars that have lost their hydrogen envelopes to form helium stars then the emission lines will reflect that and it will be more likely that a type Ib/c will occur. In the reverse situation, fewer stars will have lost their hydrogen envelope, so fewer hot helium stars contribute to the stellar population, and type II SNe become more common than type Ibc. 

These helium stars are a common prediction of stellar models that include binary interactions. Although they have not yet been observed in our own Galaxy we can be sure from their impact, in the ionizing photon production in galaxies \citep{2017A&A...608A..11G,2016MNRAS.456..485S} and in creating type Ib/c SNe, that they must exist \citep{2013MNRAS.436..774E,2016MNRAS.461L.117E}. These stars, while currently ``invisible'' nearby in our Galaxy, have important impacts on more distant stellar populations. Ignoring them can lead to inconsistencies and errors in studies of stellar systems, especially in young populations where the binary fraction is high. 

\section{Conclusion}

Our main conclusions of this work are as follows,
\begin{enumerate}
\item To accurately understand the emission line nebulae at SN sites we must account for interacting binaries in the stellar populations and allow for some loss of ionizing photons. 
\item When we derive our best-fitting ages for SN progenitors we find assuming binary stellar populations that both type II and type Ibc SN progenitor typically have ages of 10-30 Myr, and therefore they are less-massive stars with initial masses ${\rm M \leq }$ 20 $ M_{\odot} $. 
\item We find no single, monotonic relation between metallicity, initial mass and eventual SN type as predicted for single stars in \citet{2003ApJ...591..288H}. Instead binary interactions must be taken into account and we find that the expected mean SN type is related to the emission nebula created by the stellar population surrounding the SN site. This is shown by the fact that the binary models provide a much improved prediction of the type II to Ibc ratio as a function of progenitor mass and metallicity than single star models.
\item It is possible to derive an approximate relation between the age of a SN progenitor and the H$\alpha$ EW, however a more accurate age can be derived by using all available emission lines.
\end{enumerate}

These findings strongly indicate that modelling the nebular emission from stellar populations including interacting binaries must be included. Using the techniques we outline here will provide a new rigorous method to understand core-collapse SNe.

\section*{ACKNOWLEDGEMENTS}
LX acknowledge the grant from the National Key R\&D Program of China (2016YFA0400702), the National Natural Science Foundation of China (No. 11673020 and No. 11421303), and the National Thousand Young Talents Program of China. LX would also like to thank the China Scholarship Council for its funding her PhD study at the University of Auckland and the travel funding and support from the University of Auckland. L.G. was supported in part by the US National Science Foundation under Grant AST-1311862. JJE acknowledges travel funding and support from the University of Auckland. The authors also wish to acknowledge the contribution of the NeSI high-performance computing facilities and the staff at the Centre for eResearch at the University of Auckland. New Zealand’s national facilities are provided by the New Zealand eScience Infrastructure (NeSI) and funded jointly by NeSI’s collaborator institutions and through the Ministry of Business, Innovation and Employment’s Infrastructure program. URL: \url{http://www.nesi.org.nz}. We also thank the reviewer and editor for the thoughtful and helpful comments that greatly improved the paper.

\cleardoublepage
\appendix
\begin{landscape}
\section{Best-fitting parameters for CCSN host H\,II regions}

 \begin{table}
 \vspace*{0.5cm}
\caption{In no-lekakge case: the best-fitting parameters, log(Age/yr), ${\rm \log(n_H/cm^{3}})$, $log(U)$, 12 + ${\rm \log(O/H)}$ for each CCSN host H\,II regions, from single-star and binary-star populations. The full table is available online. \label{tab:bestfits}}
  \begin{tabular}{cc|cccc|cccc}
    SN name & SN type & \multicolumn{4}{c|}{Single-star models} & \multicolumn{4}{c}{Binary-star models} \\
       & & \hspace{0.5cm} ${\rm \log(Age/yr)}$ & ${\rm \log(n_H/cm^{3})}$ & ${\rm \log(U)}$ & 12 + ${\rm \log(O/H)}$ & \hspace{0.5cm} ${\rm \log(Age/yr)}$ & ${\rm \log(n_H/cm^{3})}$ & ${\rm \log(U)}$ & 12 + ${\rm \log(O/H)}$ \\
    \hline  \hline
    2000da	&	    II	&	$	7	\pm	0	$	&	$	3	\pm	0	$	&	$	-1.77	\pm	0.27	$	&	$	8.41	\pm	0.08	$	&	$	7.6	\pm	0	$	&	$	1.26	\pm	0.86	$	&	$	-3.59	\pm	0.11	$	&	$	8.77	\pm	0.03	$	\rule{0pt}{2ex}\\
   2016eob	&	    II	&	$	7.1	\pm	0.21	$	&	$	3	\pm	0.11	$	&	$	-1.5	\pm	0.13	$	&	$	7.91	\pm	0.27	$	&	$	7.8	\pm	0.03	$	&	$	1.96	\pm	0.95	$	&	$	-3.41	\pm	0.1	$	&	$	8.72	\pm	0.1	$	\rule{0pt}{2ex}\\
    2003ld	&	    II	&	$	7	\pm	0.22	$	&	$	3	\pm	0.66	$	&	$	-1.51	\pm	0.26	$	&	$	8.21	\pm	0.22	$	&	$	7.42	\pm	0.05	$	&	$	2.24	\pm	1.02	$	&	$	-3.44	\pm	0.12	$	&	$	8.85	\pm	0.26	$	\rule{0pt}{2ex}\\
    2005db	&	   IIn	&	$	8	\pm	0.22	$	&	$	3	\pm	0.11	$	&	$	-3.9	\pm	0.5	$	&	$	9.13	\pm	0.26	$	&	$	7.4	\pm	0	$	&	$	0.22	\pm	0.88	$	&	$	-3.78	\pm	0.14	$	&	$	8.77	\pm	0.15	$	\rule{0pt}{2ex}\\
    1999ge	&	    II	&	$	7.6	\pm	0.05	$	&	$	3	\pm	0.2	$	&	$	-1.63	\pm	0.16	$	&	$	8.07	\pm	0.23	$	&	$	7.98	\pm	0.05	$	&	$	2.91	\pm	0.23	$	&	$	-3.75	\pm	0.3	$	&	$	8.87	\pm	0.08	$	\rule{0pt}{2ex}\\
   2017fqo	&	    II	&	$	8	\pm	0.54	$	&	$	3	\pm	0	$	&	$	-1.5	\pm	0.72	$	&	$	8.52	\pm	0.26	$	&	$	7.76	\pm	0.05	$	&	$	2.92	\pm	0.57	$	&	$	-3.48	\pm	0.1	$	&	$	8.39	\pm	0.33	$	\rule{0pt}{2ex}\\
    2006ee	&	   IIP	&	$	7.9	\pm	0.04	$	&	$	3	\pm	0.45	$	&	$	-1.5	\pm	0.24	$	&	$	9.13	\pm	0.24	$	&	$	8	\pm	0	$	&	$	3	\pm	0.22	$	&	$	-3.64	\pm	0.2	$	&	$	8.4	\pm	0.06	$	\rule{0pt}{2ex}\\
    2011aq	&	    II	&	$	7	\pm	0	$	&	$	3	\pm	0.24	$	&	$	-1.61	\pm	0.3	$	&	$	8.39	\pm	0.03	$	&	$	7.5	\pm	0	$	&	$	1.14	\pm	0.81	$	&	$	-3.35	\pm	0.1	$	&	$	8.78	\pm	0.06	$	\rule{0pt}{2ex}\\
    2008ie	&	   IIb	&	$	7.6	\pm	0.05	$	&	$	3	\pm	0.2	$	&	$	-1.58	\pm	0.11	$	&	$	8.63	\pm	0.3	$	&	$	8	\pm	0	$	&	$	2.93	\pm	0.33	$	&	$	-3	\pm	0.2	$	&	$	8.4	\pm	0.05	$	\rule{0pt}{2ex}\\
    2009ie	&	   IIP	&	$	6.9	\pm	0.24	$	&	$	3	\pm	0.23	$	&	$	-2.69	\pm	0.33	$	&	$	8.52	\pm	0.05	$	&	$	7.5	\pm	0	$	&	$	0.83	\pm	0.76	$	&	$	-3.36	\pm	0.05	$	&	$	8.5	\pm	0.1	$	\rule{0pt}{2ex}\\

  \hline
  
  \end{tabular}  
  \end{table}
  
\begin{table}
\vspace*{0.5cm}
\caption{In lekakge case: the best-fitting parameters, log(Age/yr), ${\rm \log(n_H/cm^{3}})$, $log(U)$, 12 + ${\rm \log(O/H)}$ for each CCSN host H\,II regions, from single-star and binary-star populations. The full table is available online. \label{tab:bestfits-leak}}
  \begin{tabular}{cc|cccc|cccc}
    SN name & SN type & \multicolumn{4}{c|}{Single-star models} & \multicolumn{4}{c}{Binary-star models} \\
       & & \hspace{0.5cm} ${\rm \log(age/yr)}$ & ${\rm \log(n_H/cm^{3})}$ & ${\rm \log(U)}$ & 12 + ${\rm \log(O/H)}$ & \hspace{0.5cm} ${\rm \log(Age/yr)}$ & ${\rm \log(n_H/cm^{3})}$ & ${\rm \log(U)}$ & 12 + ${\rm \log(O/H)}$  \\
    \hline  \hline
    
    2000da	&	    II	&	$	6.28	\pm	0.27	$	&	$	2.92	\pm	0.40	$	&	$	-3.55	\pm	0.07	$	&	$	8.75	\pm	0.08	$	&	$	7.17	\pm	0.15	$	&	$	1.12	\pm	0.85	$	&	$	-3.63	\pm	0.06	$	&	$	8.93	\pm	0.00	$	\rule{0pt}{2ex}\\
   2016eob	&	    II	&	$	6.35	\pm	0.25	$	&	$	2.73	\pm	0.29	$	&	$	-3.34	\pm	0.08	$	&	$	8.58	\pm	0.10	$	&	$	7.23	\pm	0.21	$	&	$	1.25	\pm	0.80	$	&	$	-3.52	\pm	0.06	$	&	$	8.79	\pm	0.06	$	\rule{0pt}{2ex}\\
    2003ld	&	    II	&	$	6.10	\pm	0.18	$	&	$	2.48	\pm	0.92	$	&	$	-3.33	\pm	0.04	$	&	$	8.47	\pm	0.09	$	&	$	7.05	\pm	0.12	$	&	$	1.14	\pm	0.73	$	&	$	-3.33	\pm	0.09	$	&	$	8.80	\pm	0.07	$	\rule{0pt}{2ex}\\
    2005db	&	   IIn	&	$	6.32	\pm	0.28	$	&	$	2.94	\pm	0.20	$	&	$	-3.61	\pm	0.09	$	&	$	8.77	\pm	0.00	$	&	$	7.15	\pm	0.17	$	&	$	1.40	\pm	1.04	$	&	$	-3.72	\pm	0.08	$	&	$	8.98	\pm	0.08	$	\rule{0pt}{2ex}\\
    1999ge	&	    II	&	$	6.63	\pm	0.56	$	&	$	3.00	\pm	0.00	$	&	$	-3.49	\pm	0.25	$	&	$	8.86	\pm	0.11	$	&	$	7.13	\pm	0.07	$	&	$	1.81	\pm	0.89	$	&	$	-3.06	\pm	0.17	$	&	$	8.93	\pm	0.00	$	\rule{0pt}{2ex}\\
   2017fqo	&	    II	&	$	6.00	\pm	0.04	$	&	$	2.98	\pm	0.22	$	&	$	-3.46	\pm	0.09	$	&	$	8.36	\pm	0.11	$	&	$	7.18	\pm	0.07	$	&	$	2.91	\pm	0.22	$	&	$	-3.63	\pm	0.07	$	&	$	8.69	\pm	0.16	$	\rule{0pt}{2ex}\\
    2006ee	&	   IIP	&	$	6.00	\pm	0.04	$	&	$	2.98	\pm	0.20	$	&	$	-3.70	\pm	0.11	$	&	$	8.34	\pm	0.10	$	&	$	7.19	\pm	0.07	$	&	$	2.97	\pm	0.15	$	&	$	-3.81	\pm	0.08	$	&	$	8.68	\pm	0.15	$	\rule{0pt}{2ex}\\
    2011aq	&	    II	&	$	6.17	\pm	0.23	$	&	$	2.89	\pm	0.27	$	&	$	-3.32	\pm	0.06	$	&	$	8.63	\pm	0.09	$	&	$	7.10	\pm	0.10	$	&	$	1.04	\pm	0.69	$	&	$	-3.32	\pm	0.07	$	&	$	8.87	\pm	0.08	$	\rule{0pt}{2ex}\\
    2008ie	&	   IIb	&	$	7.20	\pm	0.78	$	&	$	2.93	\pm	0.18	$	&	$	-2.44	\pm	0.77	$	&	$	8.99	\pm	0.17	$	&	$	7.41	\pm	0.29	$	&	$	2.48	\pm	0.33	$	&	$	-3.40	\pm	0.25	$	&	$	8.93	\pm	0.00	$	\rule{0pt}{2ex}\\
    2009ie	&	   IIP	&	$	6.03	\pm	0.07	$	&	$	2.39	\pm	0.90	$	&	$	-3.34	\pm	0.07	$	&	$	8.39	\pm	0.05	$	&	$	7.10	\pm	0.06	$	&	$	0.85	\pm	0.75	$	&	$	-3.40	\pm	0.08	$	&	$	8.77	\pm	0.00	$	\rule{0pt}{2ex}\\
    
  \hline
  
  \end{tabular}
  
  \end{table}
  \end{landscape}

%\appendix

\section{CCSNe Ages by varying HII region definition}\label{sec:CCSNeAges}

In the main text, we identify the SN parent HII region as the stellar population present within a ${\rm 1\,kpc^2 }$ projected region centred on the explosion site. This has the advantage of scaling with the distance (or redshift, and hence projected angular size) to the source. However, the IFU data from PISCO allows alternate definition of the SNe host population.

Here, we complement the SNe age results as derived in the main text with estimates drawn from two differently defined samples. The first defines the SN host region as the observed flux detected within a fixed 1 ${\rm arcsec^2}$ aperture centered at the SN location, which better approximates a fixed slit-width method from conventional spectroscopy, but results in a physical size that depends on the redshift. The second sample identifies the nearest HII region to the explosion site and extracts the flux associated with it, rather than centring on the explosion site.

We apply the same method discussed in sections 3 and 4 to these two alternately derived samples in order to estimate the SNe ages from the properties of their nebular emission. Figures B1 and B2 show the age distribution for regions within 1 ${\rm arcsec^2}$ and for the nearest HII region sample separately. We obtain similar results as those presented in the main text in Figure 15, which suggests that our main results and their interpretation are not highly sensitive to the definition of the host population, as long as this is sufficiently representative of stars near the explosion site.

%The spectra of these CCSN hosts from PISCO can be selected depending on position relative to the SNe explosion site and nebula regions. So to some extend the so-called SN hosts are more accurately to be these SN parent HII regions. The sample we use in the main text is with their observed flux derived within 1 ${\rm kpc^{2}}$ centered at the SN location. Therefore,  the scale of it coverage does not depend on redshift. Here, we complement the SNe age results as discussed above from other two sample. The fist sample has their observed flux derived within 1 ${\rm arcsec^{2}}$ centered at the SN location, and thus this measurement can be more close to the SNe sites but depending on the redshift. Compared between the two methods above the nearest HII regions to the SNe sites are figured out. 

%We apply the similar method to these two observed sample as discussed in the previous sections to estimate the SNe ages from the properties of their nebular emission. Figures~\ref{hist_SN2} and \ref{hist_SNe3} show the age distribution of one within 1 ${\rm arcsec^{2}}$ and the nearest HII region samples separately. We obtain similar results as in Figure~\ref{hist_SN1} which can indicate that the derived spectra of the three sample are from the nebular regions of similar ionizing properties and similar underneath stellar populations. 

\begin{figure*}
\centering
%\vspace*{-0.5cm}
\textbf{Flux from HII regions within 1 ${\rm arcsec^{2}}$}\\
\vspace*{0.5cm}
\text{H$ \alpha $ EW as age indicator}\\
\includegraphics[width=8.5cm]{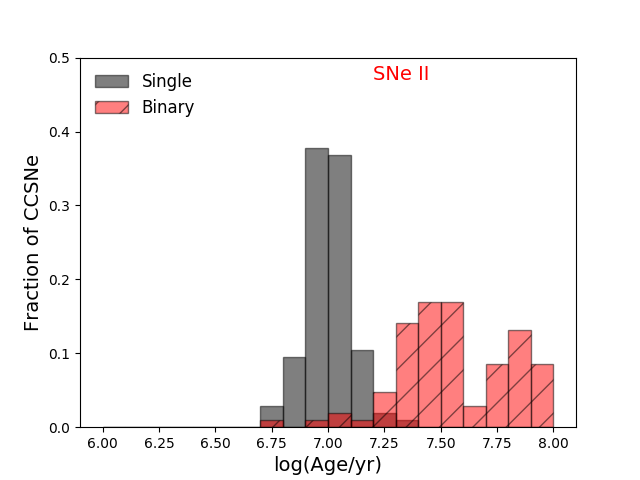}
\includegraphics[width=8.5cm]{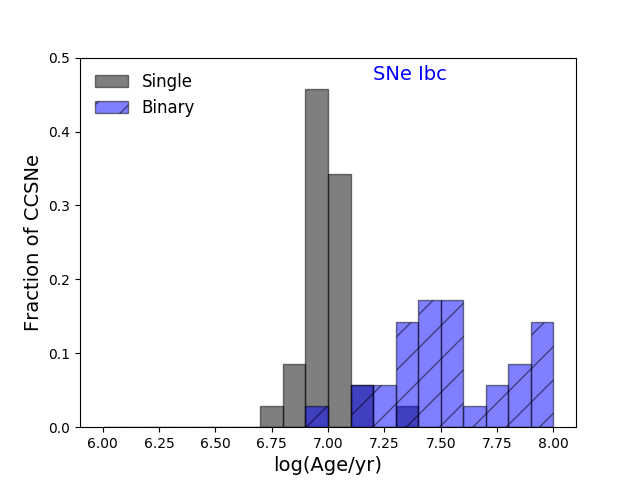}\\
\text{Age estimation in no-leakage case}\\
\includegraphics[width=8.5cm]{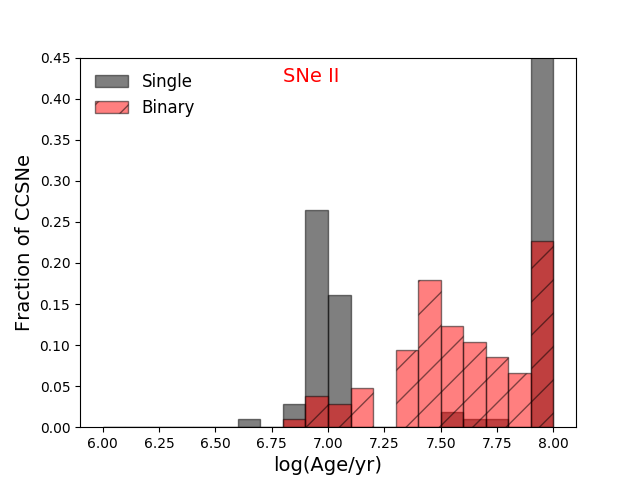}
\includegraphics[width=8.5cm]{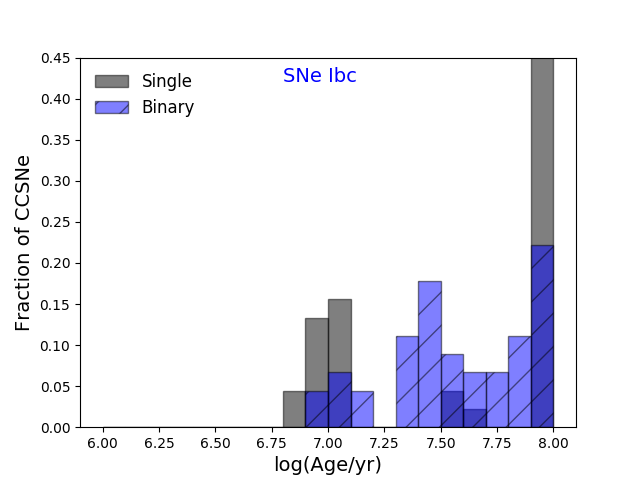}\\
\text{Age estimation in leakage case}\\
\includegraphics[width=8.5cm]{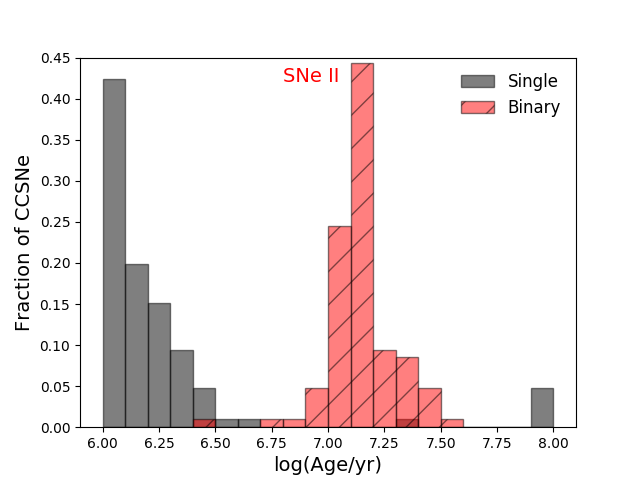}
\includegraphics[width=8.5cm]{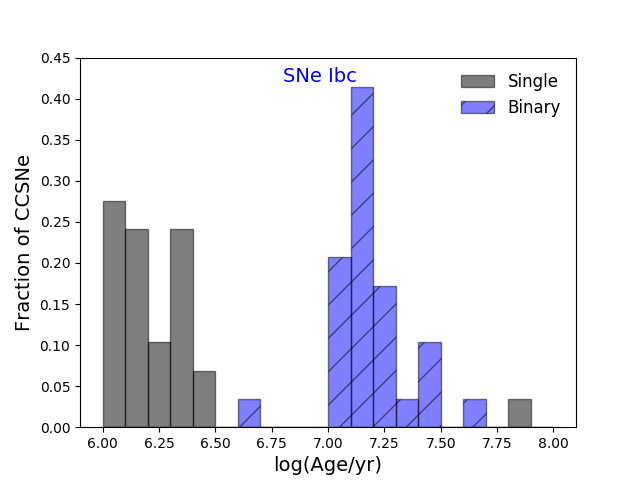}
\centering 
\caption{Age distributions derived from three methods for the sample of flux within 1 ${\rm arcsec^{2}}$: directly from H$ \alpha $ EW, best-fitting models without ionizing photon leakage and best-fitting models including leakage. The black bars are for single-star models and red (SNe II) and blue pattern (SNe Ibc) bars are for binary-star models. }\label{hist_SN2}
\end{figure*}

\begin{figure*}
\centering
%\vspace*{-0.5cm}
\textbf{Flux from the nearest HII regions}\\
\vspace*{0.5cm}
\text{H$ \alpha $ EW as age indicator}\\
\includegraphics[width=8.5cm]{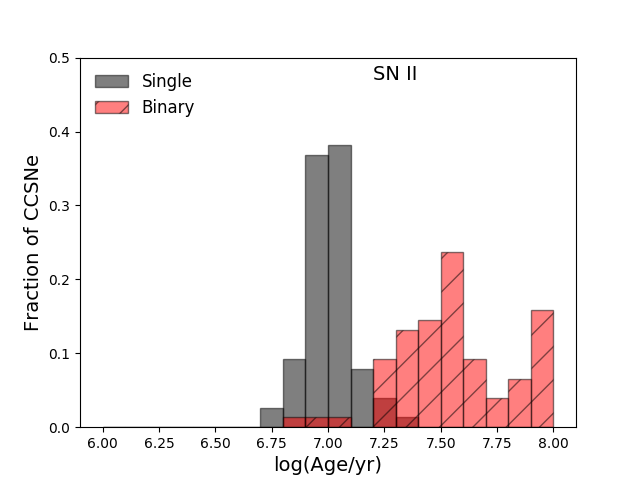}
\includegraphics[width=8.5cm]{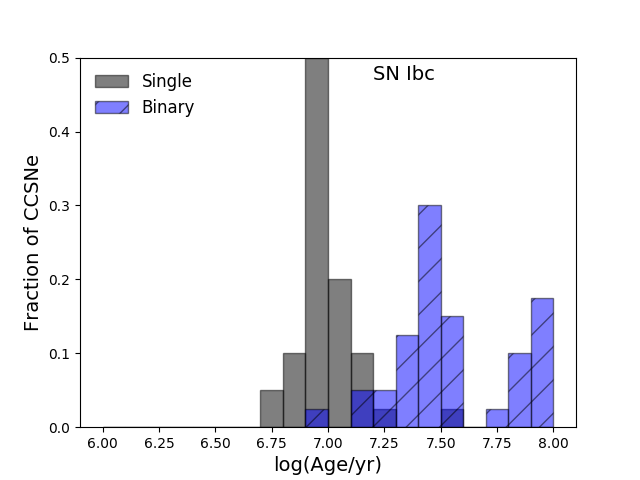}\\
\text{Age estimation in no-leakage case}\\
\includegraphics[width=8.5cm]{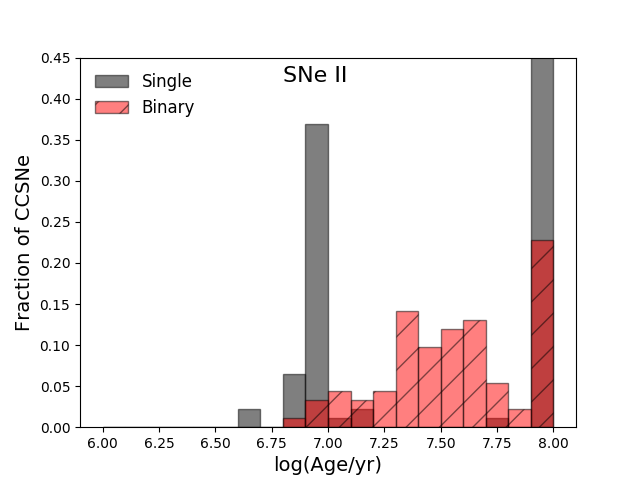}
\includegraphics[width=8.5cm]{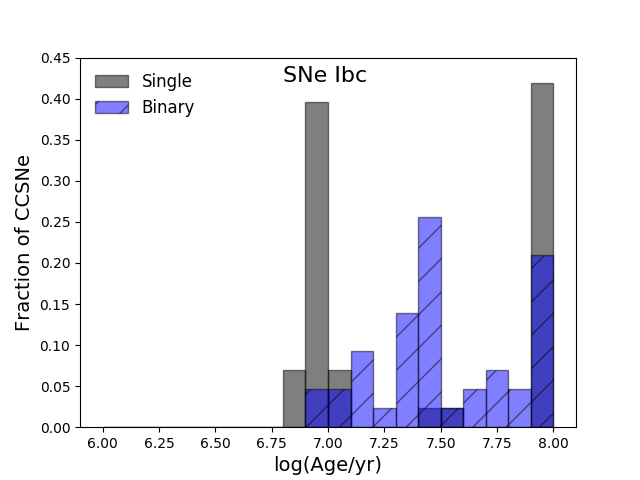}\\
\text{Age estimation in leakage case}\\
\includegraphics[width=8.5cm]{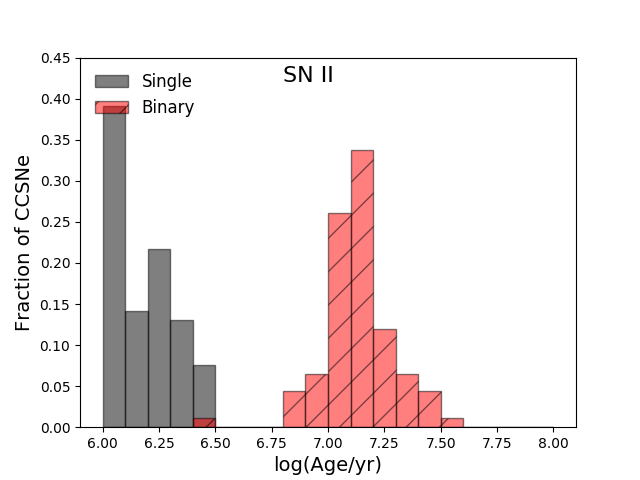}
\includegraphics[width=8.5cm]{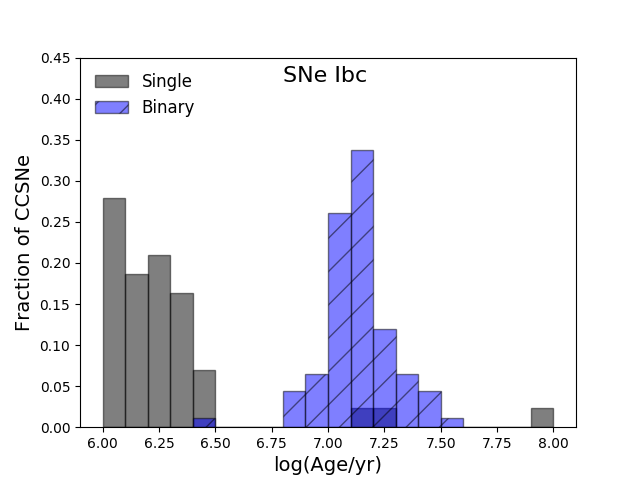}
\caption{Age distributions derived from three methods for the sample of flux from the nearest HII regions: directly from H$ \alpha $ EW, best-fitting models without ionizing photon leakage and best-fitting models including leakage. The black bars are for single-star models and red (SNe II) and blue pattern (SNe Ibc) bars are for binary-star models. }\label{hist_SNe3}
\end{figure*}

\bsp

\end{document}